\documentclass{aa}  

\usepackage{graphicx}
\usepackage{txfonts}
\usepackage{color}
\usepackage{amsmath}	
\usepackage{lscape}
\usepackage{hyperref} 
\usepackage[switch]{lineno}
\let\oldequation\equation
\let\oldendequation\endequation
\renewenvironment{equation}
{\linenomathNonumbers\oldequation}
{\oldendequation\endlinenomath}
\usepackage{subfiles} 
\begin{document} 
\title{Millisecond pulsars phenomenology \\ under the light of graph theory}
   \titlerunning{MSPs phenomenology and graph theory}
    \authorrunning{García et al.}
    \author{C. R. Garc\'{i}a\inst{1,2} \thanks{E-mail: crodriguez@ice.csic.es}, 
G. Illiano\inst{3,4,5},
D. F. Torres\inst{1,2,6}, 
A. Papitto\inst{3}, 
F. Coti Zelati\inst{1,2},
D. de Martino\inst{7},
A. Patruno\inst{1,2}
}

\def\DDM#1{{\textcolor{blue}{#1}}}
\def\ch#1{{\textcolor{green}{#1}}}
\def\red#1{\textcolor{red}{ #1}}

\institute{Institute of Space Sciences (ICE, CSIC), Campus UAB, Carrer de Can Magrans s/n, 08193 Barcelona, Spain
\and 
Institut d’Estudis Espacials de Catalunya (IEEC), E-08034 Barcelona, Spain
\and
INAF-Osservatorio Astronomico di Roma, Via Frascati 33, I-00078, Monte Porzio Catone (RM), Italy
\and
Tor Vergata University of Rome, Via della Ricerca Scientifica 1, I-00133 Roma, Italy
\and
Sapienza Università di Roma, Piazzale Aldo Moro 5, I-00185 Rome, Italy
\and
Institució Catalana de Recerca i Estudis Avançats (ICREA), E-08010 Barcelona, Spain
\and
INAF-Osservatorio Astronomico di Capodimonte, Salita Moiariello 16, I-80131 Naples, Italy
   } 

  \abstract{
We compute and apply the minimum spanning tree (MST) of the binary millisecond pulsar population, and discuss aspects of the known phenomenology of these systems in this context. 
We find that the MST effectively separates different classes of spider pulsars, eclipsing radio pulsars in tight binary systems either with a companion with a mass in the range $\sim$0.1–0.8 M$_\odot$ (redbacks) or with a 
$\lesssim0.06$ M$_\odot$ (black widows),
into distinct branches. The MST also separates black widows located in globular clusters from those found in the field and groups other pulsar classes of interest, including transitional millisecond pulsars. 
Using the MST and a defined ranking for similarity, we identify possible candidates likely to belong to these pulsar classes.
In particular, based on this approach, we propose the black widows' classification of J1300+1240, J1630+3550, J1317-0157, J1221-0633, J1627+3219, J1737-0314A, and J1701-3006F, discuss that of J1908+2105, and analyze J1723-2837, J1431-4715, and J1902-5105 as possible transitional systems.
We introduce an algorithm that quickly locates where new pulsars fall within the MST and use this to examine the positions of the transitional millisecond pulsar IGR J18245-2452 (PSR J1824-2452I), the transitional millisecond pulsar candidate 3FGL J1544.6-1125, and the accreting millisecond X-ray pulsar SAX J1808.4-3658.
Assessing the positions of these sources in the MST assuming a range for their unknown variables 
(e.g., the spin period derivative of PSR J1824-2452I) we can effectively narrow down the parameter space necessary for searching and determining key pulsar parameters through targeted observations.
}

   \maketitle

\section{Introduction}
\label{intro}

This article continues the study that integrates principal component analysis (PCA, e.g., \citealt{Pearson1901, Shlens2014}) and the application of graph theory \citep[e.g.][]{Wilson2010} to the field of pulsar astrophysics (see, e.g., \citealt{Maritz2016, MST-1, MST-2, Vohl2023}).

PCA is a dimensionality reduction technique suitable for pinpointing the variables that contain most of the variance of a sample
(see, e.g., \citealt{PCA_application}).
Graph theory provides a robust theoretical framework whose objects, the graphs, represent pulsars and their variables.
Here, we specifically consider the population of millisecond pulsars (MSPs), i.e. weakly magnetized and rotating neutron stars with spinning periods shorter than 10 ms. They are usually hosted in tight binary systems in a low mass ($< 1 \, \mathrm{M_\odot}$) companion star.
MSPs offer unique insights into stellar evolution, the interaction between magnetic fields and plasma transferred by the donor star, and particle acceleration from compact objects, particularly in binary systems (see, e.g., \citealt{Manchester2017, Papitto2022} for reviews).

By treating each MSP as a node we shall compute the MST (see e.g., \citealt{Kruskal1956, Gower1969}) of the MSP population, and use it both, to describe the population as a whole, and to identify individual pulsars that may warrant further investigation due to their unique attributes or positions within the graph.
We shall specifically examine locations in the MST in which spider pulsars reside (see, e.g., \citealt{Eichler1988, Roberts2013, Roberts20172018, DiSalvo2023}),
i.e., eclipsing radio pulsars in tight binary systems either with a non-degenerate main-sequence companion with a mass in the range $\sim$0.1–0.8 M$_\odot$ (redbacks, RBs) or with a $\lesssim0.06$ M$_\odot$ semi-degenerate companion (black widows, BWs).
We also consider transitional millisecond pulsars (tMSPs, \citealt{Papitto_tMSPSectionBook}), that exhibit dramatic state changes, i.e., go from rotation-powered, where they behave as RBs, to accretion-powered and vice versa, on timescales as short as a few weeks. 
Driven by the method developed by \cite{MST-2}, which allows for dividing an MST into distinct parts based on specific variables, we explore how to classify MSPs into different groups. 
Finally, we will use the algorithms described below to search for the positions of pulsars within the graph, specifically focusing on those whose properties have not been measured, to predict ranges for specific variables and their characteristic phenomenology.

\section{The sample variance and the MST}
\label{sample_variables}

\subsection{Sample selection}
\label{sec: sample}

The sample used is taken from the most recent version of the Australia Telescope National Facility (ATNF) catalog, v2.1.1 \citep{ATNF-Catalog} imposing that pulsars have a spin period in the millisecond range ($P<10^{-2}$ s) and a positive spin period derivative ($\dot P>0 \ \, \mathrm{s \, s^{-1}}$).
The latter allows us to calculate the intrinsic variables derived from $P$ and $\dot P$ as described in \S \ref{sec: variables_PCA}.
A total of 218 pulsars result from these cuts.
Of these, 43 are confirmed as BWs, 9 of them in globular clusters, (see eg., \citealt{swihart_blacwidows, Linares2023, Freire2017, Lynch_2012, Douglas2022}), 16 as RBs, 2 of them in globular clusters, (see e.g., \citealt{strader_redbacks, Linares2023}) and 2 as tMSPs, J1023+0038 \citep{Archibald2009} and J1227-4853 \citep{Bassa2014}.
Note that tMSP J1824-2452I \citep{IGRJ1824_INTEGRAL} is excluded from the sample as it does not have a $\dot P$ measurement, although it will be analyzed in \S \ref{sec: IGRJ1824}.

These pulsars are listed in Table \ref{tab: BW_RW_tMSP}.
According to the third pulsar catalog by the Fermi Large Area Telescope (\textit{Fermi-LAT}, see \citealt{Fermi3PC}), 
103 pulsars out of the 218 are found to be gamma-ray emitters.

\begin{table*}
	\centering
  \scriptsize
\caption{The truncated values according to v2.1.1 of the ATNF catalog (visit \url{https://www.atnf.csiro.au/research/pulsar/psrcat/} for updated versions)
of the 10 variables considered for the confirmed BWs, RBs, and tMSPs (candidates for BWs or RBs are placed below the corresponding line).
The tMSPs are included in the RBs section and are denoted by an asterisk.
Pulsars belonging to globular clusters are marked with a star.
Note that the pulsar J1300+1240 (also known as PSR B1257+12, see \citealt{Wolszczan_1992, Wolszczan_1994, Wolszczan_2000}) is a planetary system, and its orbital parameters can be considered as lower limits depending on the mass of its planets (see, e.g., \citealt{Konacki_2002}).
}
\begin{tabular}{lrrrrrrrrrrr}
\hline
      JNAME & $P_{B}(\mathrm{d})$ &  $\dot{E}_{sd}(\mathrm{erg \, s^{-1}})$ &  $P(\mathrm{s})$ &  $B_{s}(\mathrm{G})$ &  $\dot{P}(\mathrm{s \,s^{-1}})$ & $M_{C}(\mathrm{M_{\odot}})$   &  $A_{1}(\mathrm{lt-s})$ &  $\Delta\Phi(\mathrm{V})$ &  $\eta_{GJ}(\mathrm{cm^{-3}})$ &  $B_{lc}(\mathrm{G})$ \\
    \hline
        \multicolumn{11}{c}{Black widows} \\ 
\hline
                    
 J0023+0923 & 0.13 & 1.58$\times 10^{34}$ & 0.0030 & 1.88$\times 10^{8}$ & 1.14$\times 10^{-20}$ & 0.018 & 0.034 &1.33$\times 10^{14}$& 4.28$\times 10^{9}$&  62401.91 \\
 J0024-7204O$^{\star}$ & 0.13 & 6.48$\times 10^{34}$ & 0.0026 & 2.86$\times 10^{8}$ & 3.03$\times 10^{-20}$ & 0.024 & 0.045 &2.69$\times 10^{14}$& 7.50$\times 10^{9}$& 145482.85 \\
 J0024-7204P$^{\star}$ & 0.14 & 5.41$\times 10^{35}$ & 0.0036 & 1.57$\times 10^{9}$ & 6.63$\times 10^{-19}$ & 0.019 & 0.038 &7.77$\times 10^{14}$& 2.98$\times 10^{10}$& 305080.04 \\
 J0024-7204R$^{\star}$ & 0.06 & 1.38$\times 10^{35}$ & 0.0034 & 7.27$\times 10^{8}$ & 1.48$\times 10^{-19}$ & 0.029 & 0.033 &3.93$\times 10^{14}$& 1.44$\times 10^{10}$& 161687.07 \\
 J0251+2606 & 0.20 & 1.82$\times 10^{34}$ & 0.0025 & 1.40$\times 10^{8}$ & 7.57$\times 10^{-21}$ & 0.027 & 0.065 &1.42$\times 10^{14}$& 3.82$\times 10^{9}$&  80162.82 \\
 J0312-0921 & 0.09 & 1.53$\times 10^{34}$ & 0.0037 & 2.73$\times 10^{8}$ & 1.97$\times 10^{-20}$ & 0.010 & 0.015 &1.30$\times 10^{14}$& 5.10$\times 10^{9}$&  50446.32 \\
 J0610-2100 & 0.28 & 8.45$\times 10^{33}$ & 0.0038 & 2.20$\times 10^{8}$ & 1.23$\times 10^{-20}$ & 0.024 & 0.073 &9.71$\times 10^{13}$& 3.95$\times 10^{9}$&  35950.28 \\
 J0636+5128 & 0.06 & 5.76$\times 10^{33}$ & 0.0028 & 1.00$\times 10^{8}$ & 3.44$\times 10^{-21}$ & 0.007 & 0.008 &8.02$\times 10^{13}$& 2.42$\times 10^{9}$&  39963.19 \\
 J0952-0607 & 0.26 & 6.66$\times 10^{34}$ & 0.0014 & 8.31$\times 10^{7}$ & 4.77$\times 10^{-21}$ & 0.022 & 0.062 &2.72$\times 10^{14}$& 4.06$\times 10^{9}$& 275776.46 \\
 J1124-3653 & 0.22 & 1.69$\times 10^{34}$ & 0.0024 & 1.21$\times 10^{8}$ & 6.01$\times 10^{-21}$ & 0.031 & 0.079 &1.37$\times 10^{14}$& 3.49$\times 10^{9}$&  81614.11 \\
 J1301+0833 & 0.27 & 6.64$\times 10^{34}$ & 0.0018 & 1.41$\times 10^{8}$ & 1.05$\times 10^{-20}$ & 0.027 & 0.078 &2.72$\times 10^{14}$& 5.29$\times 10^{9}$& 211158.03 \\
 J1311-3430 & 0.06 & 4.93$\times 10^{34}$ & 0.0025 & 2.34$\times 10^{8}$ & 2.09$\times 10^{-20}$ & 0.009 & 0.010 &2.34$\times 10^{14}$& 6.33$\times 10^{9}$& 130950.18 \\
 J1446-4701 & 0.27 & 3.66$\times 10^{34}$ & 0.0021 & 1.48$\times 10^{8}$ & 9.80$\times 10^{-21}$ & 0.021 & 0.064 &2.02$\times 10^{14}$& 4.67$\times 10^{9}$& 131662.94 \\
 J1513-2550 & 0.17 & 8.96$\times 10^{34}$ & 0.0021 & 2.16$\times 10^{8}$ & 2.16$\times 10^{-20}$ & 0.018 & 0.040 &3.16$\times 10^{14}$& 7.06$\times 10^{9}$& 213347.04 \\
J1518+0204C$^{\star}$ & 0.08 & 6.71$\times 10^{34}$ & 0.0024 & 2.57$\times 10^{8}$ & 2.60$\times 10^{-20}$ & 0.043 & 0.057 &2.73$\times 10^{14}$& 7.17$\times 10^{9}$& 157520.28 \\
 J1544+4937 & 0.12 & 1.09$\times 10^{34}$ & 0.0021 & 7.86$\times 10^{7}$ & 2.79$\times 10^{-21}$ & 0.019 & 0.032 &1.10$\times 10^{14}$& 2.51$\times 10^{9}$&  73224.85 \\
 J1555-2908 & 0.23 & 3.07$\times 10^{35}$ & 0.0017 & 2.85$\times 10^{8}$ & 4.45$\times 10^{-20}$ & 0.059 & 0.151 &5.86$\times 10^{14}$& 1.10$\times 10^{10}$& 468636.58 \\
J1641+3627E$^{\star}$ & 0.11 & 4.47$\times 10^{34}$ & 0.0024 & 2.10$\times 10^{8}$ & 1.74$\times 10^{-20}$ & 0.023 & 0.037 &2.23$\times 10^{14}$& 5.86$\times 10^{9}$& 128450.77 \\
 J1641+8049 & 0.09 & 4.27$\times 10^{34}$ & 0.0020 & 1.36$\times 10^{8}$ & 8.94$\times 10^{-21}$ & 0.046 & 0.064 &2.18$\times 10^{14}$& 4.65$\times 10^{9}$& 154502.68 \\
 J1653-0158 & 0.05 & 1.24$\times 10^{34}$ & 0.0019 & 6.95$\times 10^{7}$ & 2.40$\times 10^{-21}$ & 0.011 & 0.010 &1.17$\times 10^{14}$& 2.44$\times 10^{9}$&  85609.20 \\
J1701-3006E$^{\star}$ & 0.15 & 3.62$\times 10^{35}$ & 0.0032 & 1.01$\times 10^{9}$ & 3.10$\times 10^{-19}$ & 0.035 & 0.070 &6.36$\times 10^{14}$& 2.16$\times 10^{10}$& 281028.45 \\
 J1719-1438 & 0.09 & 1.63$\times 10^{33}$ & 0.0057 & 2.18$\times 10^{8}$ & 8.04$\times 10^{-21}$ & 0.001 & 0.001 &4.27$\times 10^{13}$& 2.60$\times 10^{9}$&  10546.33 \\
 J1731-1847 & 0.31 & 7.78$\times 10^{34}$ & 0.0023 & 2.46$\times 10^{8}$ & 2.54$\times 10^{-20}$ & 0.038 & 0.120 &2.94$\times 10^{14}$& 7.28$\times 10^{9}$& 179657.07 \\
 J1745+1017 & 0.73 & 5.77$\times 10^{33}$ & 0.0026 & 8.60$\times 10^{7}$ & 2.72$\times 10^{-21}$ & 0.015 & 0.088 &8.03$\times 10^{13}$& 2.24$\times 10^{9}$&  43265.82 \\
 J1805+0615 & 0.33 & 9.31$\times 10^{34}$ & 0.0021 & 2.22$\times 10^{8}$ & 2.27$\times 10^{-20}$ & 0.026 & 0.087 &3.22$\times 10^{14}$& 7.23$\times 10^{9}$& 216420.69 \\
 J1810+1744 & 0.15 & 3.97$\times 10^{34}$ & 0.0016 & 8.84$\times 10^{7}$ & 4.60$\times 10^{-21}$ & 0.049 & 0.095 &2.10$\times 10^{14}$& 3.68$\times 10^{9}$& 181229.94 \\
J1824-2452M$^{\star}$ & 0.24 & 4.42$\times 10^{34}$ & 0.0047 & 7.75$\times 10^{8}$ & 1.22$\times 10^{-19}$ & 0.012 & 0.032 &2.22$\times 10^{14}$& 1.12$\times 10^{10}$&  66406.37 \\
J1824-2452N$^{\star}$ & 0.19 & 1.66$\times 10^{35}$ & 0.0033 & 7.39$\times 10^{8}$ & 1.59$\times 10^{-19}$ & 0.021 & 0.049 &4.31$\times 10^{14}$& 1.52$\times 10^{10}$& 183891.70 \\
 J1833-3840 & 0.90 & 1.07$\times 10^{35}$ & 0.0018 & 1.84$\times 10^{8}$ & 1.77$\times 10^{-20}$ & 0.009 & 0.061 &3.46$\times 10^{14}$& 6.82$\times 10^{9}$& 265593.33 \\
J1836-2354A$^{\star}$ & 0.20 & 2.42$\times 10^{33}$ & 0.0033 & 8.92$\times 10^{7}$ & 2.31$\times 10^{-21}$ & 0.019 & 0.046 &5.20$\times 10^{13}$& 1.84$\times 10^{9}$&  22164.68 \\
 J1928+1245 & 0.13 & 2.40$\times 10^{34}$ & 0.0030 & 2.27$\times 10^{8}$ & 1.67$\times 10^{-20}$ & 0.010 & 0.018 &1.63$\times 10^{14}$& 5.21$\times 10^{9}$&  77477.09 \\
 J1959+2048 & 0.38 & 1.60$\times 10^{35}$ & 0.0016 & 1.66$\times 10^{8}$ & 1.68$\times 10^{-20}$ & 0.024 & 0.089 &4.23$\times 10^{14}$& 7.16$\times 10^{9}$& 375949.48 \\
 J2017-1614 & 0.09 & 7.80$\times 10^{33}$ & 0.0023 & 7.61$\times 10^{7}$ & 2.45$\times 10^{-21}$ & 0.030 & 0.043 &9.33$\times 10^{13}$& 2.27$\times 10^{9}$&  57631.66 \\
 J2047+1053 & 0.12 & 1.04$\times 10^{34}$ & 0.0042 & 3.02$\times 10^{8}$ & 2.08$\times 10^{-20}$ & 0.040 & 0.069 &1.07$\times 10^{14}$& 4.87$\times 10^{9}$&  35979.39 \\
 J2051-0827 & 0.09 & 5.48$\times 10^{33}$ & 0.0045 & 2.42$\times 10^{8}$ & 1.27$\times 10^{-20}$ & 0.030 & 0.045 &7.82$\times 10^{13}$& 3.72$\times 10^{9}$&  24801.53 \\
 J2052+1219 & 0.11 & 3.38$\times 10^{34}$ & 0.0019 & 1.16$\times 10^{8}$ & 6.70$\times 10^{-21}$ & 0.038 & 0.061 &1.94$\times 10^{14}$& 4.06$\times 10^{9}$& 139873.60 \\
 J2055+3829 & 0.12 & 4.32$\times 10^{33}$ & 0.0020 & 4.62$\times 10^{7}$ & 9.99$\times 10^{-22}$ & 0.025 & 0.045 &6.95$\times 10^{13}$& 1.53$\times 10^{9}$&  47537.98 \\
 J2115+5448 & 0.13 & 1.67$\times 10^{35}$ & 0.0026 & 4.46$\times 10^{8}$ & 7.49$\times 10^{-20}$ & 0.024 & 0.044 &4.32$\times 10^{14}$& 1.18$\times 10^{10}$& 237535.29 \\
 J2214+3000 & 0.41 & 1.91$\times 10^{34}$ & 0.0031 & 2.16$\times 10^{8}$ & 1.47$\times 10^{-20}$ & 0.015 & 0.059 &1.46$\times 10^{14}$& 4.81$\times 10^{9}$&  67001.37 \\
 J2234+0944 & 0.41 & 1.66$\times 10^{34}$ & 0.0036 & 2.73$\times 10^{8}$ & 2.01$\times 10^{-20}$ & 0.017 & 0.068 &1.36$\times 10^{14}$& 5.21$\times 10^{9}$&  53684.04 \\
 J2241-5236 & 0.14 & 2.60$\times 10^{34}$ & 0.0021 & 1.24$\times 10^{8}$ & 6.89$\times 10^{-21}$ & 0.013 & 0.025 &1.70$\times 10^{14}$& 3.93$\times 10^{9}$& 111420.40 \\
 J2256-1024 & 0.21 & 3.71$\times 10^{34}$ & 0.0022 & 1.63$\times 10^{8}$ & 1.13$\times 10^{-20}$ & 0.034 & 0.082 &2.03$\times 10^{14}$& 4.92$\times 10^{9}$& 126750.83 \\
 J2322-2650 & 0.32 & 5.54$\times 10^{32}$ & 0.0034 & 4.54$\times 10^{7}$ & 5.83$\times 10^{-22}$ & 0.0008 & 0.002 &2.48$\times 10^{13}$& 9.08$\times 10^{8}$&  10266.96 \\
 \hline
 J1221-0633	&0.38	& 2.87$\times 10^{34}$&	0.0019&	1.02$\times 10^{8}$&	5.26$\times 10^{-21}$& 0.015&    0.05&	1.79$\times 10^{14}$&3.65$\times 10^{9}$&	132280.06\\
 J1300+1240$^\dag$   &25.2   & 1.87$\times 10^{34}$& 0.0062&	8.53$\times 10^{8}$&	1.14$\times 10^{-19}$& 5.6$\times 10^{-8}$&  3$\times 10^{-6}$&1.44$\times 10^{14}$&9.49$\times 10^{9}$&	 33265.14\\
 J1317-0157	&0.08	& 8.82$\times 10^{33}$&	0.0029&	1.27$\times 10^{8}$&	5.49$\times 10^{-21}$& 0.020&    0.02&	9.92$\times 10^{13}$&3.04$\times 10^{9}$&	 48767.61\\
 J1627+3219	&0.16	& 2.07$\times 10^{34}$&	0.0021&	1.10$\times 10^{8}$&	5.47$\times 10^{-21}$& 0.025&    0.05&	1.52$\times 10^{14}$&3.50$\times 10^{9}$&	 99738.68\\
 J1630+3550	&0.31	& 2.44$\times 10^{34}$&	0.0032&	2.62$\times 10^{8}$&	2.08$\times 10^{-20}$& 0.011&    0.03&	1.65$\times 10^{14}$&5.62$\times 10^{9}$&	 73150.90\\
 J1701-3006F$^{\star}$ &0.20	& 7.25$\times 10^{35}$&	0.0022&	7.22$\times 10^{8}$&	2.22$\times 10^{-19}$& 0.024&    0.05&	9.00$\times 10^{14}$&2.17$\times 10^{10}$&	560490.08\\
 J1737-0314A$^{\star}$ &0.22	& 4.86$\times 10^{35}$&	0.0019&	4.40$\times 10^{8}$&	9.55$\times 10^{-20}$& 0.018&    0.04&	7.36$\times 10^{14}$&1.53$\times 10^{10}$&	531633.70\\
 
    \hline
        \multicolumn{11}{c}{Redbacks and transitional millisecond pulsars} \\
\hline
 J1023+0038$^{\ast}$& 0.19 & 5.68$\times 10^{34}$ & 0.0016 & 1.09 $\times 10^{8}$ & 6.92 $\times 10^{-21}$ & 0.15 & 0.34 &2.52$\times 10^{14}$&4.48$\times 10^{9}$& 213328.66 \\
 J1048+2339 & 0.25 & 1.16$\times 10^{34}$ & 0.0046 & 3.79 $\times 10^{8}$ & 3.00 $\times 10^{-20}$ & 0.35 & 0.83 &1.14$\times 10^{14}$&5.62$\times 10^{9}$&  35000.56 \\
 J1227-4853$^{\ast}$& 0.28 & 9.12$\times 10^{34}$ & 0.0016 & 1.38 $\times 10^{8}$ & 1.10 $\times 10^{-20}$ & 0.24 & 0.66 &3.19$\times 10^{14}$&5.67$\times 10^{9}$& 270534.19 \\
 J1431-4715 & 0.44 & 6.83$\times 10^{34}$ & 0.0020 & 1.70 $\times 10^{8}$ & 1.41 $\times 10^{-20}$ & 0.14 & 0.55 &2.76$\times 10^{14}$&5.86$\times 10^{9}$& 196263.68 \\
 J1622-0315 & 0.16 & 7.93$\times 10^{33}$ & 0.0038 & 2.12 $\times 10^{8}$ & 1.14 $\times 10^{-20}$ & 0.11 & 0.21 &9.41$\times 10^{13}$&3.81$\times 10^{9}$&  34977.90 \\
 J1628-3205 & 0.21 & 1.42$\times 10^{34}$ & 0.0032 & 1.98 $\times 10^{8}$ & 1.19 $\times 10^{-20}$ & 0.18 & 0.41 &1.26$\times 10^{14}$&4.27$\times 10^{9}$&  56116.06 \\
J1717+4308A$^{\star}$ & 0.20 & 7.65$\times 10^{34}$ & 0.0031 & 4.44 $\times 10^{8}$ & 6.11 $\times 10^{-20}$ & 0.18 & 0.39 &2.92$\times 10^{14}$&9.73$\times 10^{9}$& 132169.23 \\
 J1723-2837 & 0.61 & 4.65$\times 10^{34}$ & 0.0018 & 1.19 $\times 10^{8}$ & 7.54 $\times 10^{-21}$ & 0.27 & 1.22 &2.28$\times 10^{14}$&4.46$\times 10^{9}$& 175618.45 \\
J1740-5340A$^{\star}$ & 1.35 & 1.36$\times 10^{35}$ & 0.0036 & 7.92 $\times 10^{8}$ & 1.68 $\times 10^{-19}$ & 0.21 & 1.65 &3.90$\times 10^{14}$&1.50$\times 10^{10}$& 152737.57 \\
 J1803-6707 & 0.38 & 7.49$\times 10^{34}$ & 0.0021 & 2.00 $\times 10^{8}$ & 1.84 $\times 10^{-20}$ & 0.33 & 1.06 &2.89$\times 10^{14}$&6.51$\times 10^{9}$& 193679.03 \\
 J1816+4510 & 0.36 & 5.22$\times 10^{34}$ & 0.0031 & 3.75 $\times 10^{8}$ & 4.31 $\times 10^{-20}$ & 0.18 & 0.59 &2.41$\times 10^{14}$&8.13$\times 10^{9}$& 108100.42 \\
 J1908+2105 & 0.14 & 3.23$\times 10^{34}$ & 0.0025 & 1.90 $\times 10^{8}$ & 1.38 $\times 10^{-20}$ & 0.06 & 0.11 &1.90$\times 10^{14}$&5.14$\times 10^{9}$& 105961.15 \\
 J1957+2516 & 0.23 & 1.74$\times 10^{34}$ & 0.0039 & 3.33 $\times 10^{8}$ & 2.74 $\times 10^{-20}$ & 0.11 & 0.28 &1.39$\times 10^{14}$&5.82$\times 10^{9}$&  50306.20 \\
 J2039-5617 & 0.22 & 3.00$\times 10^{34}$ & 0.0026 & 1.96 $\times 10^{8}$ & 1.41 $\times 10^{-20}$ & 0.19 & 0.47 &1.83$\times 10^{14}$&5.11$\times 10^{9}$&  98678.12 \\
 J2215+5135 & 0.17 & 7.41$\times 10^{34}$ & 0.0026 & 2.98 $\times 10^{8}$ & 3.33 $\times 10^{-20}$ & 0.24 & 0.46 &2.87$\times 10^{14}$&7.91$\times 10^{9}$& 157526.43 \\
 J2339-0533 & 0.19 & 2.32$\times 10^{34}$ & 0.0028 & 2.04 $\times 10^{8}$ & 1.41 $\times 10^{-20}$ & 0.30 & 0.61 &1.60$\times 10^{14}$&4.89$\times 10^{9}$&  79742.09 \\
  \hline
 J1902-5105*& 2.01 & 6.86$\times 10^{34}$ & 0.0017 & 1.28$\times 10^{8}$ & 9.19$\times 10^{-21}$& 0.18 & 1.90 & 2.76$\times 10^{14}$ & 5.08$\times 10^{9}$ & 227053.37 \\
 J1302-3258 & 0.78 & 4.81$\times 10^{33}$ & 0.0037 & 1.58$\times 10^{8}$ & 6.54$\times 10^{-21}$& 0.17 & 0.92 & 7.33$\times 10^{13}$ & 2.91$\times 10^{9}$ &  27786.94 \\
 \hline
\hline
\end{tabular}
\label{tab: BW_RW_tMSP}
\end{table*}

\subsection{Variables and PCA}
\label{sec: variables_PCA}

\begin{figure*}
  \includegraphics[width=0.85\textwidth]{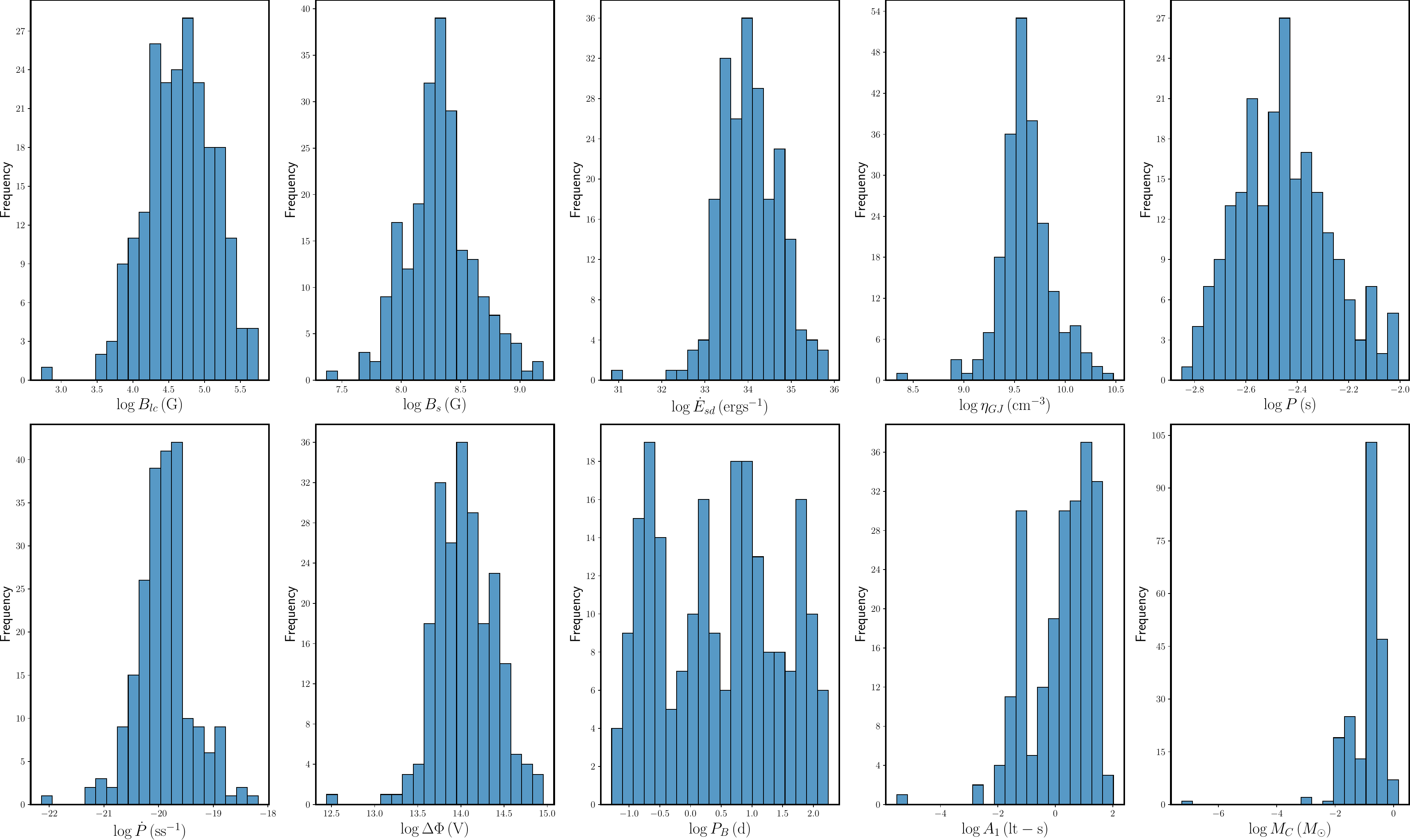}
  \centering
  \caption{Distribution of the logarithm of the 10 variables considered for the sample of 218 pulsars.   }
  \label{distr_var}
\end{figure*}

\begin{figure*}
  \includegraphics[width=1\textwidth]{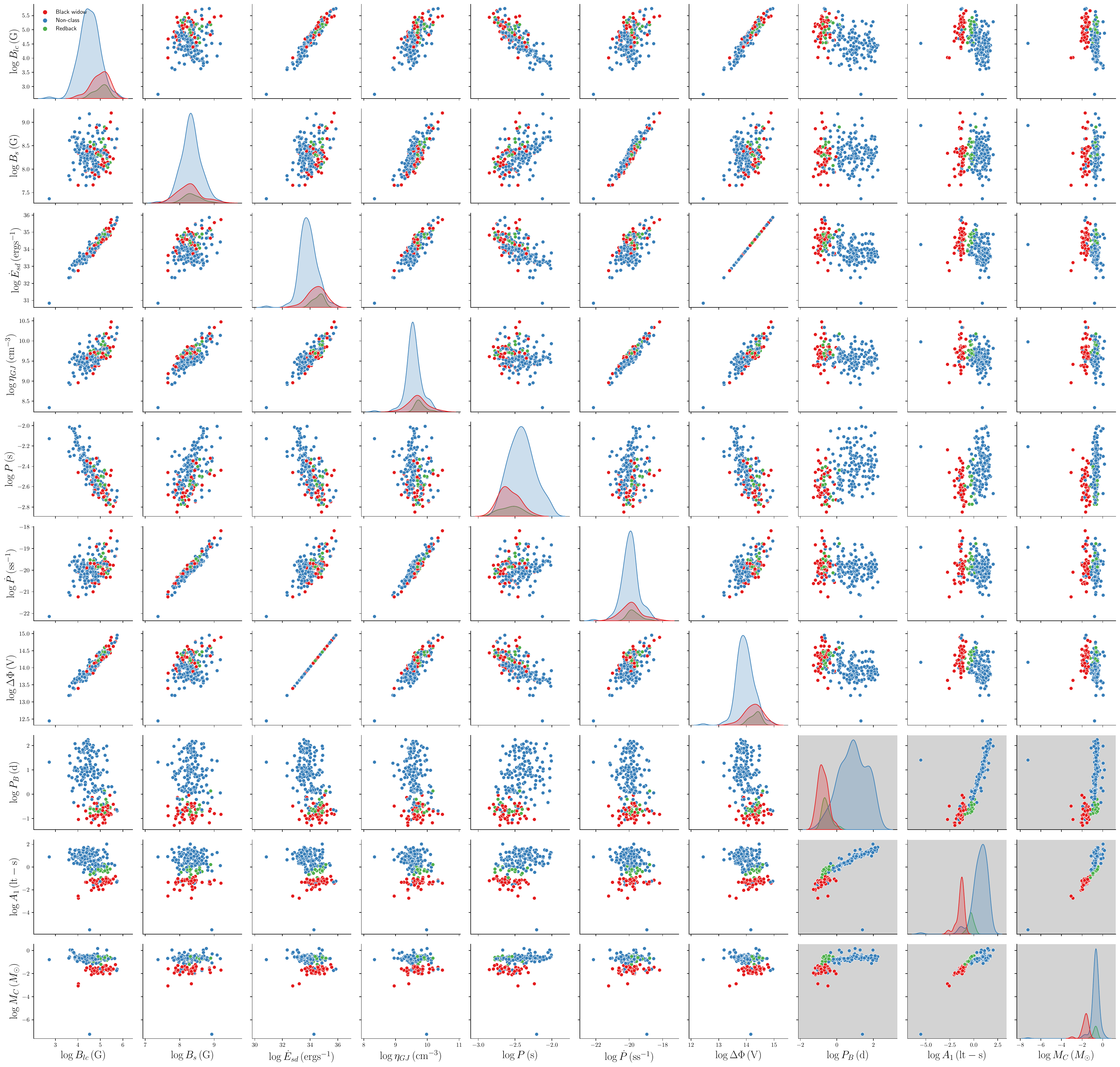}
  \centering
\caption{
Cross-dependence of the 10 variables considered.
BWs (in red) and RBs (in green) confirmed are separately noted.
For visualization reasons, no distinction is made between MSPs residing in globular clusters, and the tMSPs are marked in green due to their behavior as RB. 
The pulsars not yet assigned to any of these classes are shown in blue. 
The main diagonal shows the distribution for each variable.
The panels with the shaded background show the pairs formed with $P_{B}$, $M_C$, and $A_{1}$, which best differentiate the BWs from the rest.
}
  \label{pair_plotclasses11v}
\end{figure*}

Figure \ref{distr_var} shows the logarithmic distribution of the spin period ($P$) and spin period derivative, ($\dot P$), surface magnetic flux density  ($B_{s}$), the magnetic field at the light cylinder ($B_{lc}$), spin-down energy loss rate ($\dot{E}_{sd}$), surface electric voltage ($\Delta \Phi$), and Goldreich-Julian charge density ($\eta_{GJ}$), binary period ($P_B$), projected semi-major axis of the orbit ($A_1$) and the median mass of the companion star for each system ($M_C$).
We do not consider here the characteristic age ($\tau_c = P/ 2 \dot P$), because, in binary systems, additional torques imparted on the pulsar during accretion phases can render this estimate unreliable (see, e.g., \citealt{Kiziltan_2010, Tauris_2012, Jian2013}).
The distributions are not normal (which is also the case for the set of distributions of the original variables, without logarithm), and due to this fact we use the robust scaler to scale them (subtract the median and divide by the interquartile range, see, e.g., \citealt{RobustScaler}).
Figure \ref{pair_plotclasses11v} shows how the $P_{B}$, $A_{1}$, and $M_C$ distribution distinguishes BWs from other classes.
However, when none of these variables are used in the analysis, the distinction is no longer evident, and spider pulsars are seen as a more homogeneous group.

\begin{figure*}
  \centering
  \includegraphics[width=0.9\textwidth]{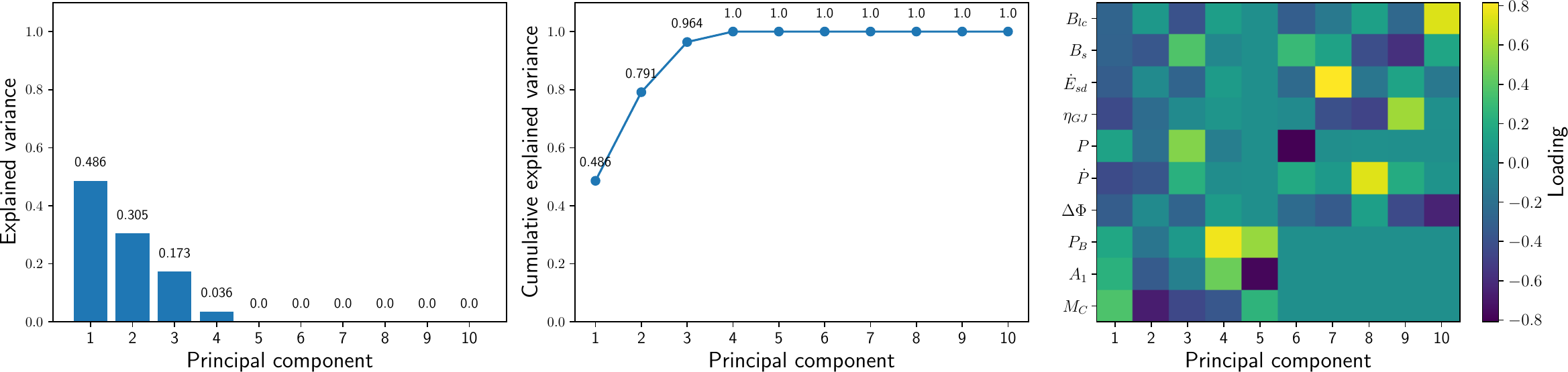}
  \caption{
  PCA results for the logarithm of the set of variables for the whole population of 218 MSPs.
  The left panel, also called the scree plot, shows the explained variance of each PC according to the covariance matrix's eigenvalues.
  Note that the covariance matrix calculates the relationships between pairs of variables, showing how changes in one variable are associated with changes in another. 
  It represents the amount of information contained in each PC.
  The central panel shows the cumulative explained variance by the new variables defined through the PCA analysis. 
  The right panel shows the 'weight', also called loading, each variable has concerning each PC, indicating its contribution to the variance captured by that PC. 
  This value is the coefficient held in each eigenvector of the covariance matrix.
  Negative values imply that the variable and the PC are negatively correlated. 
  Conversely, a positive value shows a positive correlation between the PC and the variable.
  }
  \label{PCA_10v}
\end{figure*}

%
Figure \ref{PCA_10v} shows the results of the PCA analysis applied to these variables.
In the left panel of Fig. \ref{PCA_10v}, we observe how most of the variance is distributed in the first 2 principal components (PCs).
Likewise, the central panel shows that the whole variance is represented in a 4-dimensional space, i.e., we need the first 4 PCs to cover 100\% of the variance of the sample.
This relatively low number of PCs results from the fact that within the dipolar model used as a proxy (see, e.g., \citealt{Lorimer2012}), all physical variables depend on $P$ and $\dot P$, and also that $A_{1}$ of a binary system is related to the $P_{B}$ and the $M_{C}$ (in this case through the total mass of the system, i.e., the sum of the masses of the pulsar and its companion) via Kepler's third law.
The PCA, as shown in the right panel of Fig. \ref{PCA_10v}, 
does not highlight the dominant influence of key variables for the separation between pulsar spiders, such as the $M_C$, but provides a more complex classification where its dominance is balanced with other variables.
Technically, the four PCs needed to describe all the variance have similarly large loadings in several variables (see the right panel of Fig. \ref{PCA_10v}).
%

\subsection{MST}
\label{sec: MST}

\begin{figure}
\centering
\includegraphics[width=0.49\textwidth]{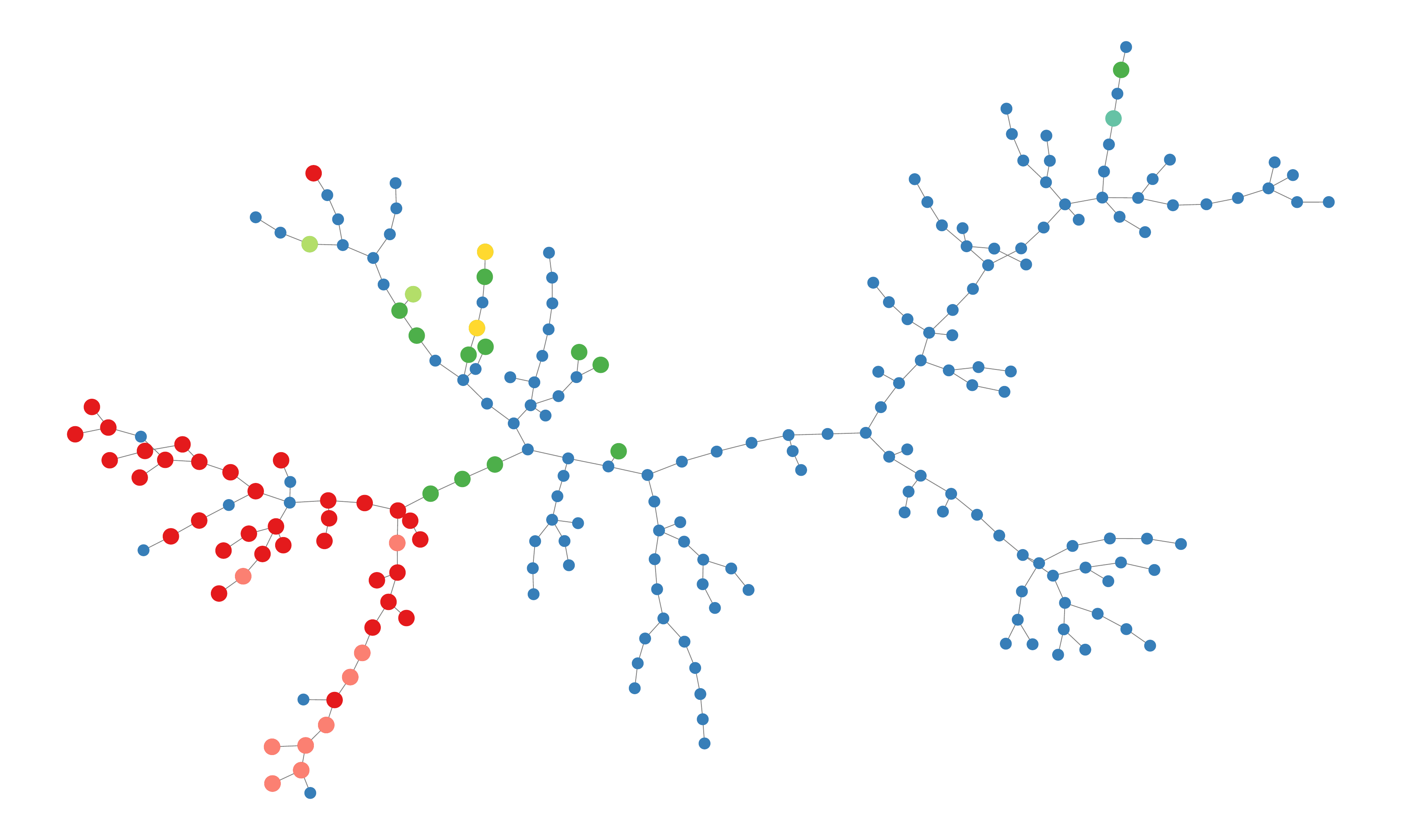}
\caption{The MST of the binary pulsar population defined as $T(218, 217)$ based on the complete, undirected, and weighted graph $G(218, 23653)$ computed from the Euclidean distance among 10 scaled variables (or the equivalent 4 PCs that described their whole variance). 
Each node in the MST represents a pulsar. 
The MST notes separately confirmed BWs (red), RBs (green), and tMSPs J1023+0038 and J1227-
4853 (yellow), respectively. 
Also, the BWs and RBs in globular clusters (light red and light green, respectively) are highlighted.
The RBs J1622-0315 (green) and the RB candidate J1302-3258 (light teal) are also noted in the rightmost branch of the MST.
The unclassified ones appear in blue.
See Table \ref{tab: BW_RW_tMSP} for more details.}
       \label{fig: MST_MSP}
\end{figure}

\cite{Kruskal1956, Kleinberg2005} and \cite{Erickson2019} explain the necessary concepts to calculate an MST. 
We refer the reader to these references for details.
We define a Euclidean distance over the variables or the principal components described above, as using the first four PCs ($\sim$ 100\% of the explained variance) produces the same MST as using all variables (and thus the results from the analysis that follows from the graph are the same) but is less demanding given the reduced dimensionality of the problem. 
With this Euclidean distance, we first compute a complete, undirected, and weighted graph $G(V, E)=G(218, 23653)$, with $|V|$\footnote{The notation $|\cdot|$ represents the cardinality (or size) of the specified set, indicating the number of elements within it.}=218 nodes and $|E|=23653$ edges, with a specific weight value defining each edge.
From that, we obtain the MST of this sample, $T(V, E')=(218, 217)$, with $|V|=218$ nodes and $|E'|=217$ edges, shown in Fig. \ref{fig: MST_MSP}.

\begin{figure*}
\includegraphics[width=1\textwidth]{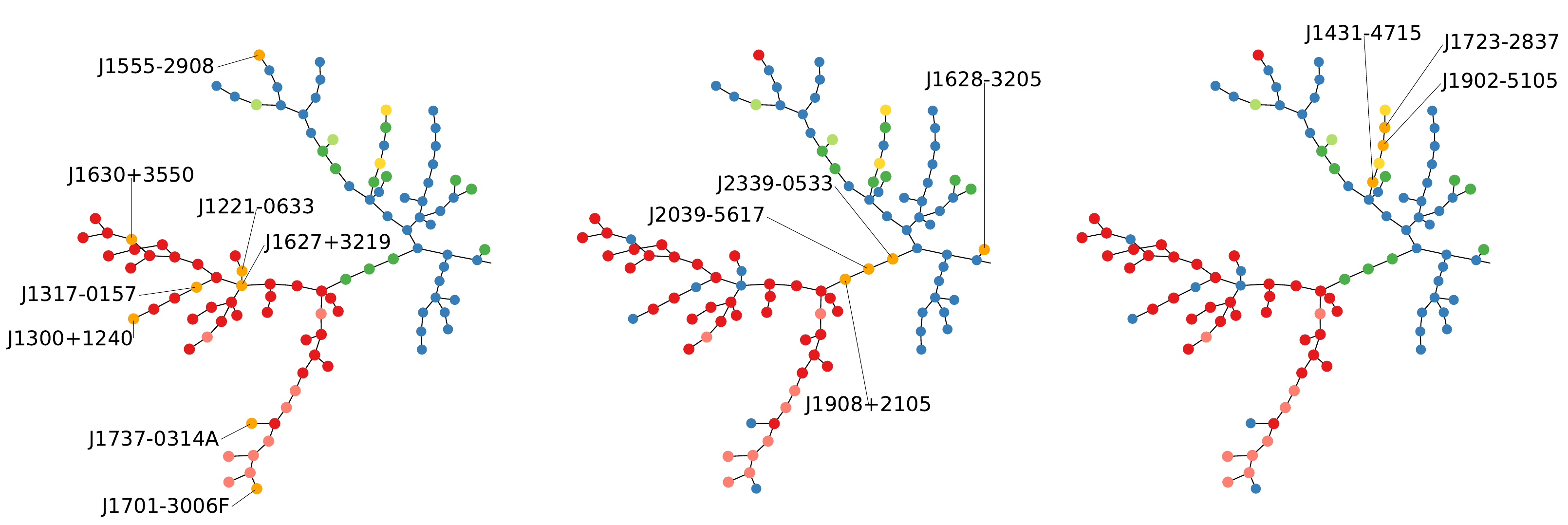}
  \centering
  \caption{
Zoom of the leftmost part of the MST shown in Fig. \ref{fig: MST_MSP}, to observe more clearly the location of the pulsars (shown in orange) discussed in \S \ref{subsection: BWs_MST} (left panel), in \S \ref{subsection: RBs_MST} (central panel), and \S \ref{subsection: tMSPs_MST} (right panel).
 The same color code of Fig. \ref{fig: MST_MSP} is shown, where the confirmed BWs appear in red, RBs in green, tMSPs in yellow, and the BWs and RBs in globular clusters in light red and light green, keeping the unclassified pulsars in blue.
 }
  \label{fig: MST_classes}
\end{figure*}

The MST can separate BWs from RBs, with tMSPs appearing close to each other, in the same structure, as shown in Fig. \ref{fig: MST_MSP}. These tree´s regions are zoomed in Fig. \ref{fig: MST_classes}, where several pulsars that will be discussed next are highlighted.
We provide an online tool\footnote{\url{http://www.pulsartree.ice.csic.es/millisecondpulsartree}} to allow the reader to zoom in, identify, and mark different portions of the MST for further study. 

%

\subsubsection{Black widows pulsars in the graph}
\label{subsection: BWs_MST}

We will now focus on some specific pulsars (see Fig. \ref{fig: MST_classes}).
Two of these are J1300+1240 and J1630+3550, which have not been explicitly classified as BWs yet but have been discussed by \cite{Yan2013, Sobey_2022}.
A possible formation scenario for J1300+1240, suggested by \cite{Yan2013}, is in low-mass, narrow-orbit bound binaries. Here, the initial point of departure is a low-mass binary pulsar with a very narrow orbit, such as the known BW J1959+2048. Because of their comparable velocity and rotation period to those of J1959+2048, J1300+1240 may evolve in this scenario.
J1630+3550 is an MSP in a binary system of 7.6h orbiting a companion with a minimum mass of 0.0098 M$_{\odot}$. 
These parameters can indicate a BW as is stated in \cite{Lewis_2023}.
From the MST graph, we can see that they are located in the BW region. 
We can further consolidate their potential categorization as BWs by observing the ranking of the nearest pulsars based on the Euclidean distance derived from the defined variables, which can be obtained using the online tool we provide with this paper.
Note that the ranking simply provides a distance-ordered list of one pulsar relative to all others, while the MST provides a global way to connect all pulsars considering all distances, optimizing the total path.
Only the nearest neighbor to any given pulsar will have an immediately recognized position within the MST, as such, the nearest neighbor will always be one of the pulsars connected to the one of interest in the MST.
For a deeper discussion and examples regarding the differences between the ranking and the MST, see \citealt{MST-1}).
J1300+1240 has 9 pulsars among its 10 nearest neighbors that have already been classified as BWs, and the remaining is J1630+3550, whose 10 nearest neighbors are BWs.

Among the interesting sources highlighted in Fig. \ref{fig: MST_classes} are J1317-0157 and J1221-0633, classified as BW systems by \cite{Swiggum_2023}, having tight orbits, low-mass companions, and exhibiting eclipses.
In both cases, their nearest neighbor ranking contains 9 confirmed BWs except for J1627+3219.

The position of J1627+3219 in the MST, underscored by its ranking, where eight of its ten nearest neighbors are confirmed BW pulsars, supports its categorization in this group. 
This assertion is further bolstered by \cite{Braglia2020}, who identified a candidate optical periodicity, necessitating verification through subsequent observations. 
Recent research by \cite{Corcoran2023} also underscores the potential BW nature of this pulsar. Still, the lack of detected radio pulsations highlights the need for further studies to understand its nature fully.

We also pinpoint J1737-0314A which, according to the timing solution provided by \cite{Zhao_2022}, is likely a BW system.
Looking at its ranking, its categorization as a BW is favored, since it contains 9 confirmed BWs among its 10 nearest neighbors, the remaining one being J1701-3006F.
In addition, the latter, J1701-3006F, considering its mass limit and orbital features, is also favored as a BW, as outlined in \cite{Freire_2005}, although eclipses have not been observed. 
However, this absence could be due to a low orbital inclination \citep{Vleeschower_2024}.
On the other hand, its nearest neighbor ranking shows six BWs with the closest being J17001-3006F, see also \cite{King2005}.
Note that both J17001-3006E and J1701-3006F appear connected in the MST (see Fig. \ref{fig: MST_classes}), filling the end of a branch that contains a large number of these cases.

The position of J1555-2908 in the MST, being the only confirmed BW away from the rest, is noteworthy.
In the work by \cite{Kennedy_2022}, extensive analysis was performed using optical spectroscopy and photometry.
This study, combined with $\gamma$-ray pulsation timing information alongside a companion mass of 0.06 M$_{\odot}$, concluded that J1555-2908 lies at the observed upper boundary of what is typically classified as a BW system.
Note that its nearest neighbor, J1835-3259B, is not a BW.

Of interest to all sources discussed in this section, we remark that no other non-BW pulsar, except J1908+2105 discussed in the next section, has more than five confirmed BWs in its ranking.

\subsubsection{Redbacks pulsars in the graph}
\label{subsection: RBs_MST}

Figure \ref{fig: MST_classes} shows that RB pulsars do not appear clustered like BWs, but are mostly close to each other, with a small fraction near the border of BWs.
The caveat here is the smaller number of RBs known so far. 

The pulsar J1908+2105 is an interesting case within the RB class.
This one is right on the frontier with the part of the MST mostly populated by BWs and could be classified as such, especially due to its minimum companion mass of 0.06 M$_{\odot}$ and its short $P_B$ of 0.14 days \citep{strader_redbacks}.
However, the pronounced radio eclipses of this system align it more closely with RBs (also see, e.g., \citealt{Linares2023, Deneva_2021}).
Note that the ranking of the 10 nearest neighbor pulsars to J1908+2105 shows that 8 out of 10 are BWs. 
The fifth and ninth pulsars in the ranking, J2039-5617 and J0337+1715 respectively, are RB and an uncategorized pulsar. 
This situation, where the ranking of a given pulsar thought to pertain to one class (RB) is dominated by pulsars belonging to the other, only arises in the case mentioned above.
Thus, the MST location and the ranking favor, or at the very least do not rule out, a BW classification for J1908+2105.

On the other hand, the gamma-ray source 3FGLJ2039-5617 (PSR J2039-5617) is almost certainly associated with an optical binary that is listed as an RB candidate by \cite{Strader_2019}. Its predicted nature of RB is validated by \cite{Corongiu_2021}, where they found clear evidence of eclipses of the radio signal for about half of the orbit which they associate with the presence of intra-binary gas. It appears as a confirmed RB in \cite{Linares2023}.
Furthermore, J2039-5617 shares similarities with the confirmed RB J2339-0533, exhibiting a peak in gamma-ray emission close to a minimum in its optical emission, which contrasts with the expected phase of the intrabinary shock (IBS) as seen in \cite{An_2020}. 
This similarity between J2039-5618 and J2339-0533 and their position in the MST highlight intriguing aspects of their behavior, alerting us to further analysis. 

The distinct positions of J1622-0315, J1302-3258 (see Fig. \ref{fig: MST_MSP} for the location of the latter two in the MST), and J1628-3205, located far from the rest of the RBs in the MST, also capture attention. 
J1622-0315 is one of the lightest known RB systems, a companion mass of 0.15 M$_{\odot}$, with a relatively hot companion \citep{yap2023light, Sen_2024}. 
This system is notable for its $P_{B}$ of 0.16 days, marginally smaller than other RBs that typically have one of 0.2 days or more. 

On the other hand, J1302–3258 is classified as an RB candidate and reported to have a minimum companion mass of 0.15 M$_{\odot}$ \citep{Strader_2019}.
However, it is distinguished by the absence of an identified optical companion and lacks published evidence of extensive radio eclipses. 
\cite{Linares2023} note that, unlike most known RBs and candidates, J1302-3258 does not have a \textit{Gaia} counterpart, a trend more typical of BWs, which generally possess cooler companion stars with fainter optical magnitudes. 
J1628-3205 is a RB in a 0.21~day-long orbit around and has a 0.17--0.24 M$_{\odot}$ companion star \citep{Ray2012, Li2014, strader_redbacks}. Its optical counterpart shows two peaks per orbital cycle. Such a shape is reminiscent of the ellipsoidal deformation of a star that nearly fills its Roche-lobe-filling star in a high-inclination binary. It suggests that heating by the pulsar wind is not a major effect. In this context, J1628-3205 is an intermediate case among RBs, falling between systems where the companion star's emission is dominated by pulsar irradiation (e.g., J1023+0038) and those that are not (e.g., J1723-2837, J1622-0315, J1431-4715)(see \citealt{Li2014,yap2023light,deMartino2024} and references therein).

\subsubsection{Transitional pulsars in the graph}
\label{subsection: tMSPs_MST}

The pulsars J1723-2837, J1902-5105, and J1431-4715 appear in the same part of the MST as the known tMSPs J1227-4853 and J1023+0038, see Fig. \ref{fig: MST_classes}.
In addition, these three pointed out are the only ones in the sample containing at least one known tMSP in their top three neighbors, see Table \ref{tab: tMSP_ranking}.
The closest neighbors for the two known tMSPs are also shown in Table \ref{tab: tMSP_ranking}.

\begin{table}
\caption{Ranking based on the Euclidean distances calculated over the 10 scaled variables considered for the nodes in the region of the 
tMSPs.}
\scriptsize
\centering
\begin{tabular}{ccccc}
\hline
            \textbf{J1023+0038} & \textbf{J1723-2837} & \textbf{J1902-5105} & \textbf{J1227-4853} & \textbf{J1431-4715} \\ 
\hline
                   J1723-2837 & J1902-5105 & J1723-2837 & J1902-5105 & J2205+6012  \\
                   J1227-4853 & J1023+0038 & J1227-4853 & J1431-4715 & J1227-4853  \\
                   J1902-5105 & J1227-4853 & J1431-4715 & J1723-2837 & J1342+2822B \\
                   
\hline
\end{tabular}
\label{tab: tMSP_ranking}
\end{table}

J1723-2837 is a nearby (d$\sim$1 kpc based on \textit{Gaia} parallax measure) 1.86~ms RB in a 15~h-period binary system \citep{Crawford2013}. With an X-ray luminosity of $10^{32}$erg s$^{-1}$, it ranks among the brightest RBs in that band \citep[see, e.g.][]{Lee2018}. Based on the similarity with the X-ray output of tMSPs in the rotation-powered pulsar state, \cite{Linares2014} suggested that it is one of the most promising candidates to observe a transition to an accretion state.
The observed emission, in both soft \citep{Bogdanov2014, Hui2014} and hard X-rays \citep{Kong2017}, is modulated at the $P_{B}$, indicative of an origin in the IBS between the pulsar wind and mass outflow from the companion star. On the other hand, its optical counterpart shows no sign of significant irradiation with two peaks per orbital cycle \citep{Li2014}.

J1902–5105, a 1.74~ms radio MSP in a 2~days-period binary system. 
It was discovered within the Parkes telescope surveys targeting unidentified \textit{Fermi-LAT} sources \citep{Kerr_2012ApJ}. It is a relatively bright MSP located at a distance of $\sim$1.2 kpc \citep{Camilo_2015ApJ}. It was suggested that the companion is a $0.2-0.3 \, \mathrm{M_{\odot}}$ white dwarf with helium core \citep{Camilo_2015ApJ}.

Finally, J1431-4715, discovered in the High Time Resolution Universe (HTRU) survey with the Parkes radio telescope is an RB MSP with a $P$ of 2.01 ms in a 10.8 hr orbit with a companion mass of 0.20 M$_{\odot}$ \citep{ bates2015mnras, MiravalZanon_2018J, deMartino2024}.

One of the most efficient ways to identify candidate tMSPs has turned out to be searching for the peculiar variability in the X-ray emission between two intensity levels (known as `high' and `low' modes) observed in all the three confirmed tMSPs in the so-called sub-luminous disc state \citep[see, e.g.,][]{Patruno2014, deMartino_2013A&A, Archibald_2015ApJ, Papitto2013, Linares_2014MNRAS, Bogdanov_2015ApJ, Papitto_2019ApJ, Baglio_2023A&A, Papitto_tMSPSectionBook}. This method has proven successful in identifying a few candidates \citep[see, e.g.,][]{CotiZelati_2019, Bogdanov_2015}.
However, a methodology for identifying candidates among the radio-emitting pulsars out of the sub-luminous disc state is unclear. The MST can contribute to this.

Although J1723-2837, J1902-5105, and J1431-4715 are currently detected as radio MSP and are therefore not expected to exhibit the typical bimodality in the X-ray light curve, we analyzed \textit{XMM-Newton} archival observations of these sources (ObsID 0653830101 for J1723-2837, ObsID 0841920101 for J1902-5105, and ObsID 0860430101 for J1431-4715) to check for possible mode switching.
We confirm the absence of moding in our analysis, see Appendix \ref{sec: appendix2}.
This implies a lack of conclusive evidence about the transitional nature of J1723-2837, J1902-5105, and J1431-4715, but 
their position in the MST and their neighbor ranking confirm them as subjects of interest as potential transitional systems, though currently in the radio pulsar state. 
The relative proximity of J1723-2837 and particularly J1902-5105 renders them optimal candidates for potentially unveiling a future transition from a rotation-powered state to an accretion-powered one.

\section{MST clustering }
\label{significance_branches}

Here we apply a clustering algorithm aimed at the separation of an MST into different parts following a quantitative prescription.
We describe the methodology in the Appendix \ref{sec: appendix1}.
Figure \ref{fig: MST_siginificance_branches} shows separated branches, and Fig. \ref{fig: siginificance_branches_Vars} shows the distributions for each magnitude considering them. 
This is how the MST visually represents the binary millisecond pulsar population, with no prior assumption on the nature of the nodes.

\begin{figure*}
\centering
\includegraphics[width=0.8\textwidth]{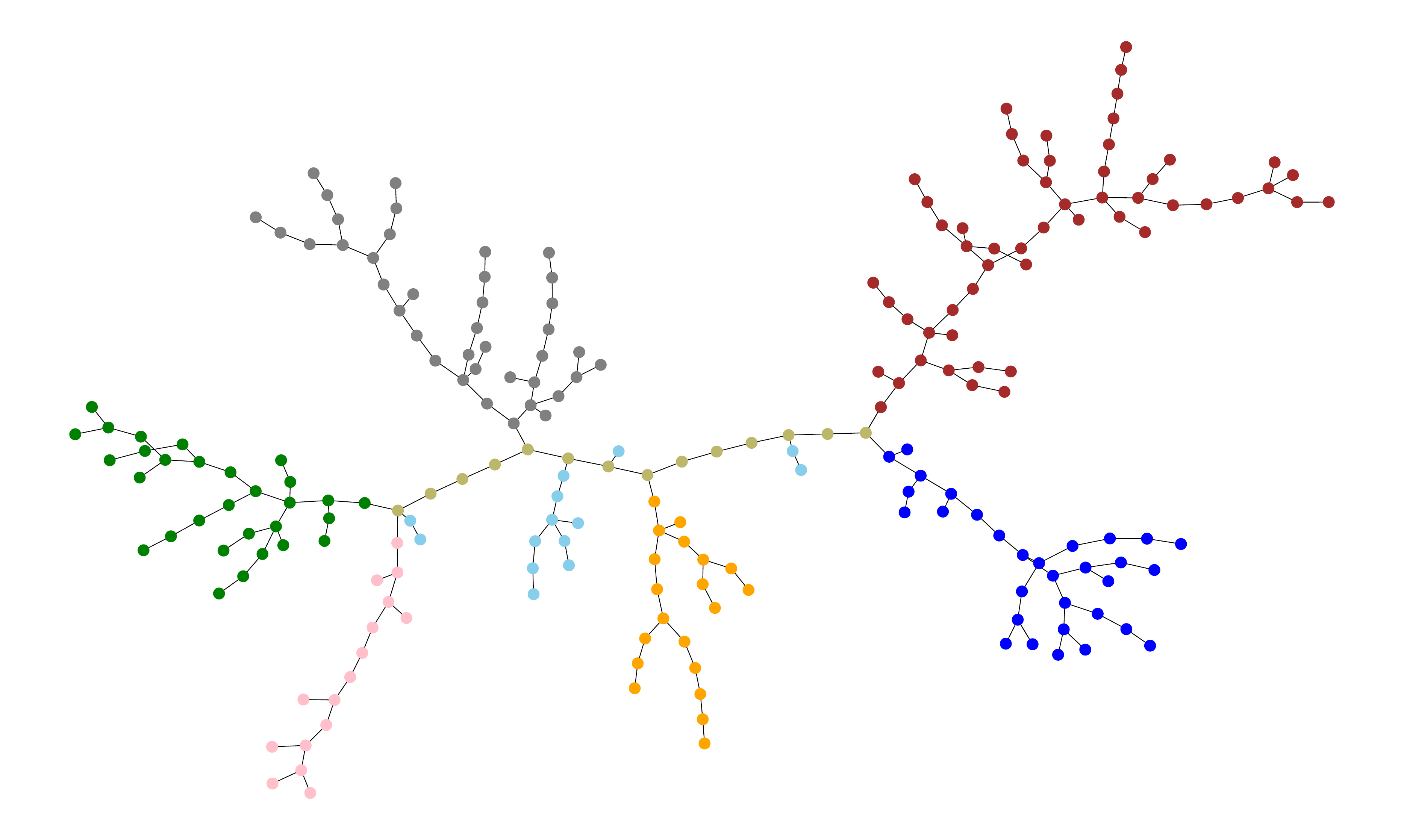}
\caption{The MST defined as $T(218, 217)$ of the MSP population is separated into significant branches according to the algorithm described in the text (see \S \ref{significance_branches}), is shown.
The branches group a comparable number of pulsars: gray (39), orange (20), dark blue (31), brown (54), dark green (30), and pink (16).
The trunk (in dark khaki) has only 14 nodes, with 14 others (in sky blue) not considered to pertain to any significant branch or trunk, being the noise of the former structures.
}
       \label{fig: MST_siginificance_branches}
\end{figure*}

\begin{figure*}
\centering
\includegraphics[width=1\textwidth]{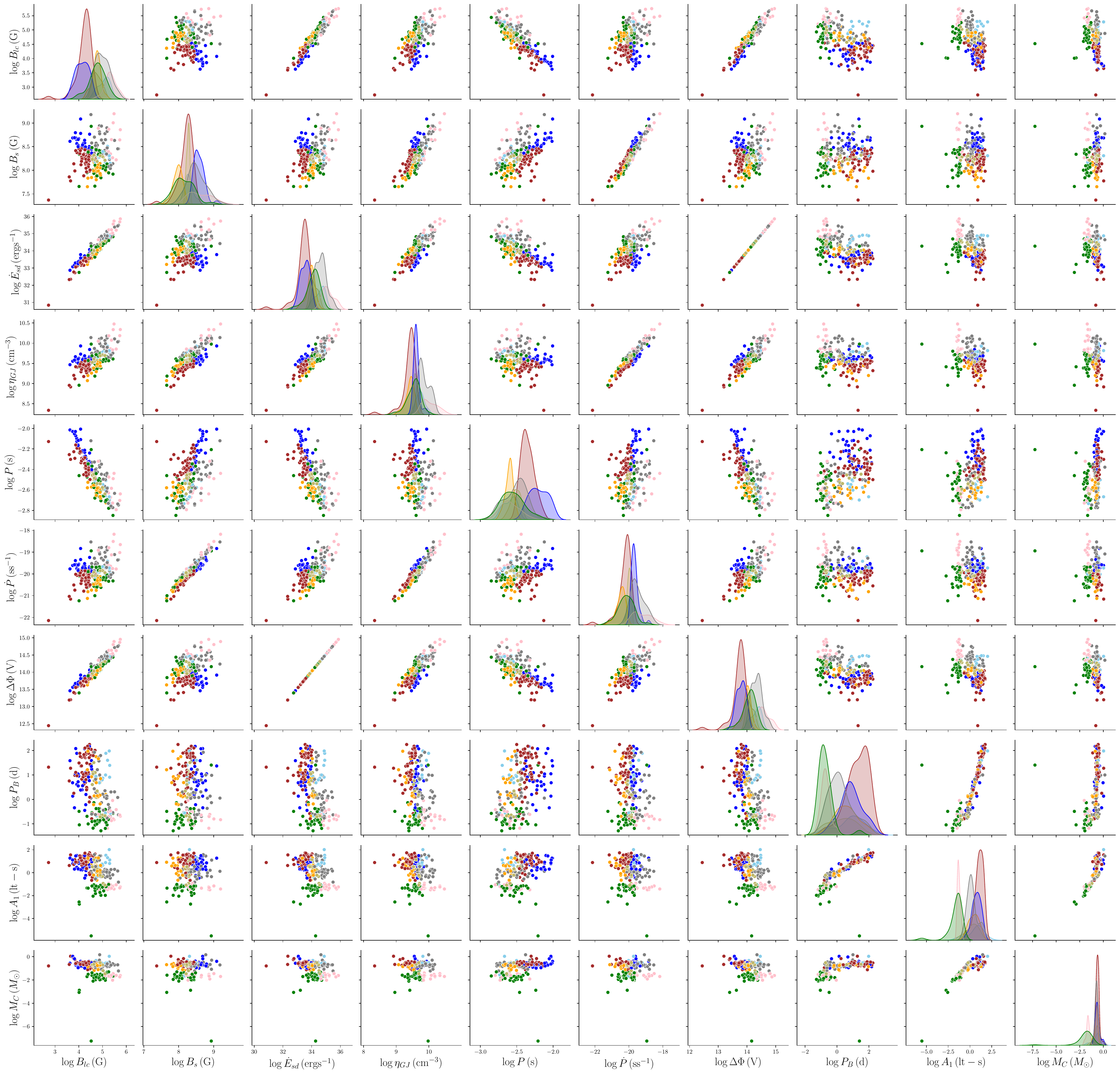}
\caption{Cross-dependence and distributions of the 10 variables considered separated according to the branches in Fig. \ref{fig: MST_siginificance_branches}. 
}
       \label{fig: siginificance_branches_Vars}
\end{figure*}

Two of the branches indicated by the method in Fig. \ref{fig: MST_siginificance_branches} (depicted in dark green and pink) contain all BWs except J1555-2908, which was already deemed as lying at the edge of the BWs' population as discussed in \S \ref{subsection: BWs_MST}.
In addition, one of these branches (pink) contains most of the candidate or confirmed BWs seen in globular clusters. 
The BWs in globular clusters together with the rest of the pulsars seen in the pink branch show a $\dot P$ and $\dot E$ in general higher than the others; see Fig. \ref{fig: siginificance_branches_Vars}.
Note that these values cannot be reliably measured in a globular cluster due to the acceleration in the cluster gravity well.
Similarly, tMSPs and most of the RBs fall in another branch (depicted in gray according to Fig. \ref{fig: MST_siginificance_branches}).
As we see in Fig. \ref{fig: siginificance_branches_Vars}, pulsars in the orange branch exhibit a $P_{B}$, $M_{C}$, and $A_{1}$ close to those in the gray branch.
However, in contrast, they show a lower $\dot{P}$ and thus a smaller $B_{s}$.
The rest of the population, particularly those nodes in the dark blue and brown branches, which are less different from each other, are markedly different from the other groups as shown in Table \ref{tab: significance_branches}.
Figure \ref{fig: siginificance_branches_Vars} shows that although these nodes are less obviously separated in the usual representations, such as the typical $(P,\dot P)$-diagram, they show larger $P_{B}$, $M_{C}$, $P$, and smaller $\dot E$ than the BW pulsars.

Based on the distributions of the variables seen in the main diagonal of Fig. \ref{fig: siginificance_branches_Vars}, we can discuss differences among branches.
We see that $M_{C}$ shows sharp distributions around similar values, except for the (pink) branch that mostly contains BWs in globular clusters, for which $M_{C}$  shows smaller values.  
The dark green branch, plagued by BWs, displays a wider distribution, always within small values.
This is reflected in $A_1$ and $P_B$, where the behavior is similar even with somewhat wider distributions, mainly for the branches that do not contain BWs. 
The widest distributions are observed in $P$, for all branches but the one depicted in orange --which has a high density of cases around low values. The dark blue branch contains the pulsars with longer periods.
The $\dot P$ shows different trends for some branches. The orange, brown, and dark green branches, in that order, show sharp behaviors that allow us to delimit more clearly the pulsars they contain. Most of them are not classified as BWs or RBs.
This is reflected in $\dot E$, $\eta_{GJ}$, $B_{lc}$, and $B_S$;  branches containing BWs and RBs show a wider distribution, shifted towards larger values for the dark green, gray, and pink branches, considering that the latter two contain different types of spider pulsars.
In addition, Fig. \ref{fig: siginificance_branches_Vars} can be compared with Fig. \ref{pair_plotclasses11v}. The former is more informative, as the clustering technique separates the nodes in groups that are significantly different from one another (see Table \ref{tab: significance_branches} in Appendix \ref{sec: appendix1}) beyond the known RBs and RBs.

\begin{figure*}
  \includegraphics[width=1\textwidth]{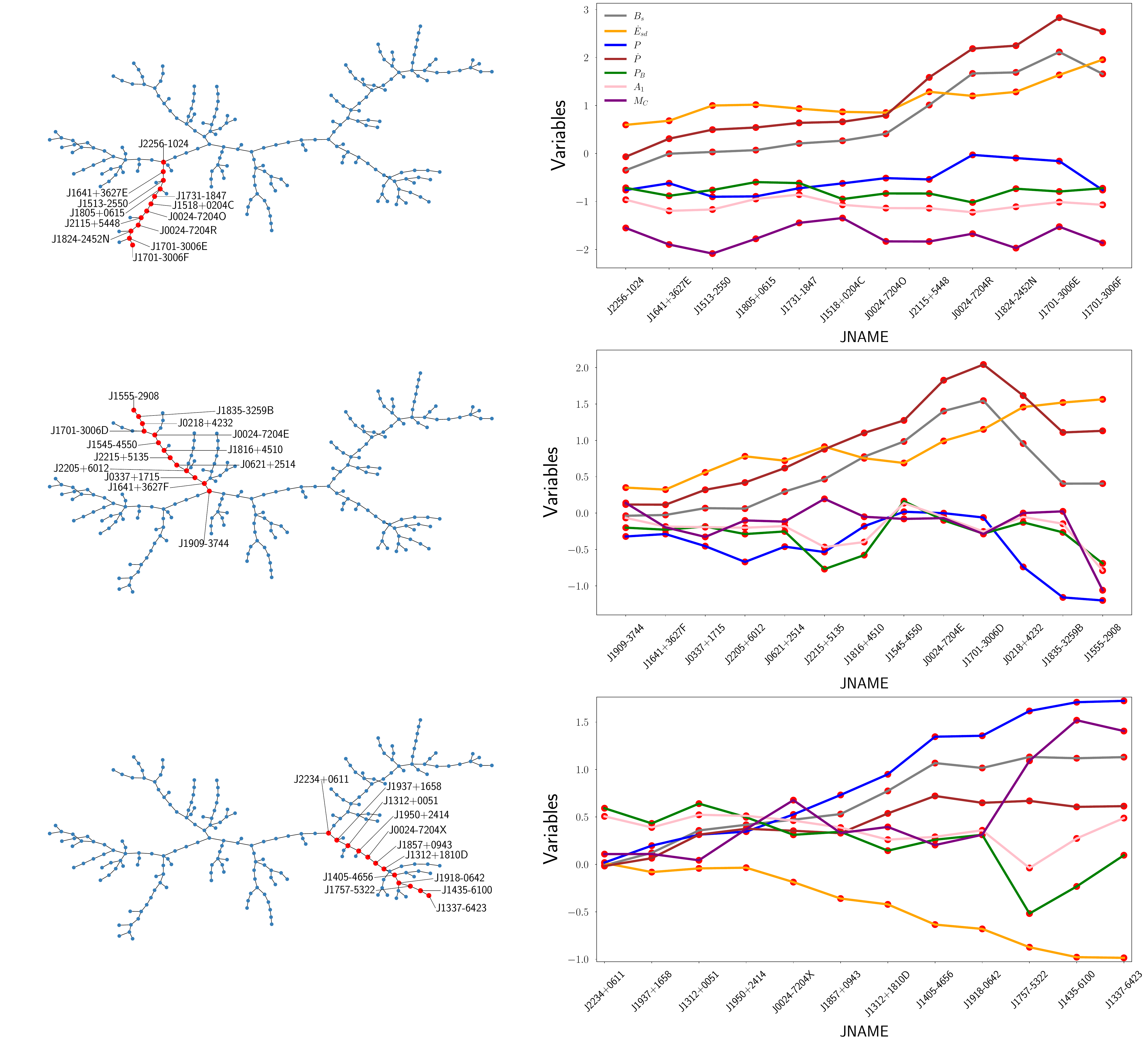}
  \centering
  \caption{
    Examples of the values of $P$, $\dot P$, $B_{s}$, $\dot E$, $M_{C}$, $A_1$, and $P_{B}$ (the legend shown in the first row applies to all rows) along a path in the MST.
    Left column: selected paths, nodes in red, in the MST are shown.
    Right column: The y-axis shows the scaled variables (using the robust scaler, i.e., subtracting the median and dividing by the interquartile range for the whole population considered). The x-axis shows the pulsars of the selected path (nodes in red) through their identifier according to the ATNF catalog. The order shown responds to the path reading in the MST, which always goes from the central part (trunk) to the ends of the MST.} 
  \label{track_variables}
\end{figure*}

In closing this section we note that the MST technique orders the pulsars according to one or several of the variables considered (this was discussed at length already in the first work by \citep{MST-1} for the whole pulsar population). Figure \ref{track_variables} below serves as an example (data to construct similar figures for other paths are provided in the online tool that accompanies this paper) of how the variables are ordered along a given sequence of consecutive nodes (or path). Figure \ref{track_variables} shows the values of $P$, $\dot P$, $B_{s}$, $\dot E$, $M_{C}$, $A_1$, and $P_{B}$ along a path in the MST.

\section{Localizing a new node}
\label{sec: predict_link}

We introduce an incremental algorithm for updating the MST when new nodes are considered. 
This can be used to infer where a new node with given properties will fall in the MST.
In practice, when we add a new node and compute its distance from all other nodes in the graph, we preserve the known distances among the previously existing nodes.
In this process, we keep the scaling of the distance used in the calculation of the original MST, denoted by $T$, to which the new node will be added. 
This involves identifying potential cycles that would be formed with existing $T$ adding new edges and discarding them so that the resulting graph remains an MST.
This approach, while avoiding the complete recalculation of the MST, delivers the location of the new node concerning the former ones, leading to an updated $T$, which we shall refer to as $T'$.
This process follows the principles outlined in traditional graph theory, e.g., the correctness of Prim's algorithm as per the MST theorem, and the MST lemma as the cycle property (see, e.g., \citealt{Prim1957, Wilson2010, Roughgarden2019}).

\subsection{IGR J18245-2452—PSR J1824-2452I}
\label{sec: IGRJ1824}

Since tMSP IGR J18245-2452 (or PSR J1824-2452I, see, e.g., \citealt{Papitto2013}) lacks a measured value of the $\dot P$, it is initially excluded from the sample used in this work (see \S \ref{sec: sample}).
Considering its measured values for $P$, $P_{B}$, $M_{C}$, and $A_{1}$, 
we shall assume a value for the $\dot P$ and add this pulsar to $T$, in Fig. \ref{fig: MST_MSP}.
Based on the estimated 
$B_{s}\sim (0.7-35) \times 10^{8}$ G by \cite{Ferrigno2014}, and its known $P$, we can compute\footnote{For consistency, we assume $B_{s} = 3.2 \times 10^{19} \sqrt{P \, \dot{P}}$ G} a range of $\dot{P}\sim (1.2\times 10^{-21} - 3\times 10^{-18}) \mathrm{ss^{-1}}$.
We take 10 equispaced values in this range (and their concurrent values of all physical variables derived taking $\dot P$ into account we construct for each one a $T'$.

\begin{figure*}
\centering
\includegraphics[width=1\textwidth]{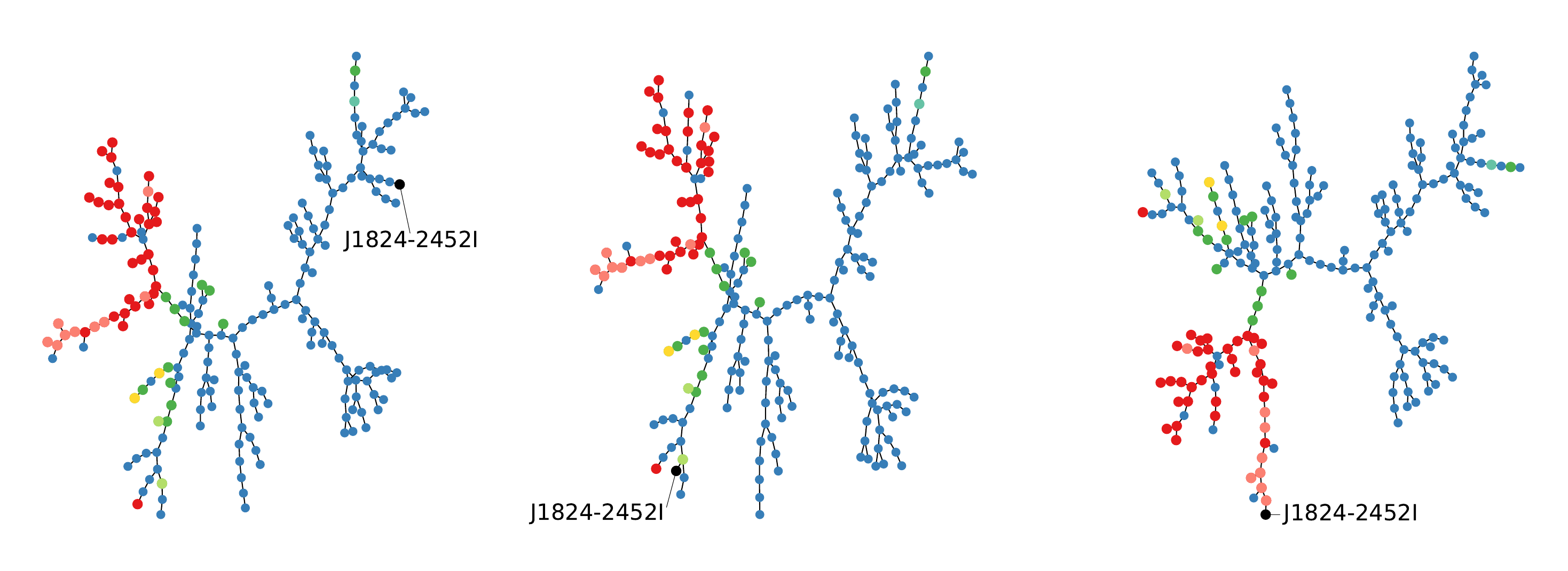}
\caption{The MSTs containing different realizations for $\dot P$, 
$\dot{P}\sim 1.22\times10^{-20}\mathrm{ss^{-1}}$ (left panel);
$\dot{P}\sim 3.34\times10^{-19}\mathrm{ss^{-1}}$ (central panel);
$\dot{P}\sim 3\times10^{-18}\mathrm{ss^{-1}}$ (right panel),
of the tMSP J1824-2452I (in black).
Each MST is defined as a $T'(219,218)$ and shown with the spider type of pulsars as seen in Fig. \ref{fig: MST_MSP}, where the confirmed BWs appear in red, RBs in green, tMSPs in yellow, the BWs and RBs in globular clusters in light red and light green, the RB candidate in light teal, and keeping the unclassified ones in blue.
}
\label{fig: MST_IGRJ181245-2452}
\end{figure*}

When we assume the lowest values of the given range, $\dot{P}\lesssim 10^{-20}\mathrm{ss^{-1}}$, we find that the added pulsar is located along one of the rightmost branches (brown branch, see Fig. \ref{fig: MST_siginificance_branches}) of the MST, see the left panel in Fig. \ref{fig: MST_IGRJ181245-2452}.
As $\dot{P}$ increases to the range ($10^{-20}, 3\times10^{-19}) \, \mathrm{ss^{-1}}$, its MST location would climb until it reaches the end of the branch where the rest of the RBs, and therefore the known tMSPs, appear. 
It will remain at that end until it $\dot{P}\lesssim 6\times 10^{-19}\mathrm{ss^{-1}}$ is considered, see the central panel in Fig. \ref{fig: MST_IGRJ181245-2452}.
When this value of $\dot P$ is exceeded and up to the largest ones explored $\dot{P}\sim 3\times 10^{-18}\mathrm{ss^{-1}}$, the pulsar would be located in the BW branch as we can see in the right panel of Fig. \ref{fig: MST_IGRJ181245-2452}.
Thus, as IGR J18245-2452 is an RB transitional pulsar, it would be reasonable to expect that it falls near the majority of the nodes of its class, implying we can limit searches for its $\dot P$ from $\dot{P}\sim (1.2\times 10^{-21} - 3\times 10^{-18})\, \mathrm{ss^{-1}}$ to just around $\dot{P}\sim (1\times 10^{-20}, 6\times 10^{-19}) \, \mathrm{ss^{-1}}$ or $B_{s}\sim (2\times 10^{8}, 1.55\times 10^{9})$G.

\subsection{tMSP candidate: 3FGL J1544.6-1125}

3FGL J1544.6-1125 has shown variability with high and low modes that are exclusively seen in tMSPs, see \cite{Bogdanov_2015}. 
In \cite{Britt_2017}, the $P_{B}$ of this pulsar is estimated to be $P_{B} = 0.2415361(36) \, \mathrm{days} = 20868.72(31) \, \mathrm{s}$. The semi-amplitude of the radial velocity of the companion star $K_2 = 39.3 \pm 1.5 \, \mathrm{km \, s^{-1}}$ yields a companion mass $M_{2}\lesssim 0.7 \, M_\odot$.
To estimate the range of $A_{1}$, we start defining the center of mass location, relying on the fundamental relationship $ M_1 a_1 = M_2 a_2$  (see e.g., \citealt{FundamentalAstronomy}), where $a_1$ and  $a_2$ represent the distances of the objects from the center of mass, and $M_1$ and $M_2$ are the masses of the neutron star and the companion star, respectively. 
Additionally, the radial velocity of the companion is defined as (see, e.g., \citealt{tauris2003formation}) $K_2 = \omega_{B} \cdot a_2 \cdot \sin(i)$, where $\omega_{B} = {2 \pi}/{P_{B}}$.
By using the above relationships, we derive \(A_{1} = a_1 \sin(i) = \frac{K_2 \cdot P_{B}}{2 \pi} \cdot \frac{M_2}{M_1} \).
We make several assumptions, where 
for $M_1$ we have
$ M_{1min}\sim 1.4 M_{\odot}$
and 
$ M_{1max}\sim 2 M_{\odot} $, 
and for $M_2$ we have
$M_{2min}\sim 0.05 M_{\odot}$
and 
$M_{2max}\sim 0.7 M_{\odot}$. 
Consequently, we calculate the range $A_1\sim(0.0088, 0.259)$ lt-s.
We consider values from $M_{2min}$ up to $M_{2max}$ for the range of $M_{C}$.
Also, as the pulsar has an unknown value of $P$, we impose no constraint on it and we shall range it within $\sim (1,10)\times10^{-3}$ s.
Similarly, we shall also inspect the range between the minimum and maximum measurement of $\dot{P}$ seen in the sample according to $(7 \times  10^{-23}, 7 \times 10^{-19})$ ss$^{-1}$. 
We consider 10 values spanning each range for ($P$, $\dot P$, $A_1$, $M_{C}$), resulting in $10^4$ distinct combinations for which we apply the incremental algorithm. 

\begin{figure}
\centering
\includegraphics[width=0.5\textwidth]{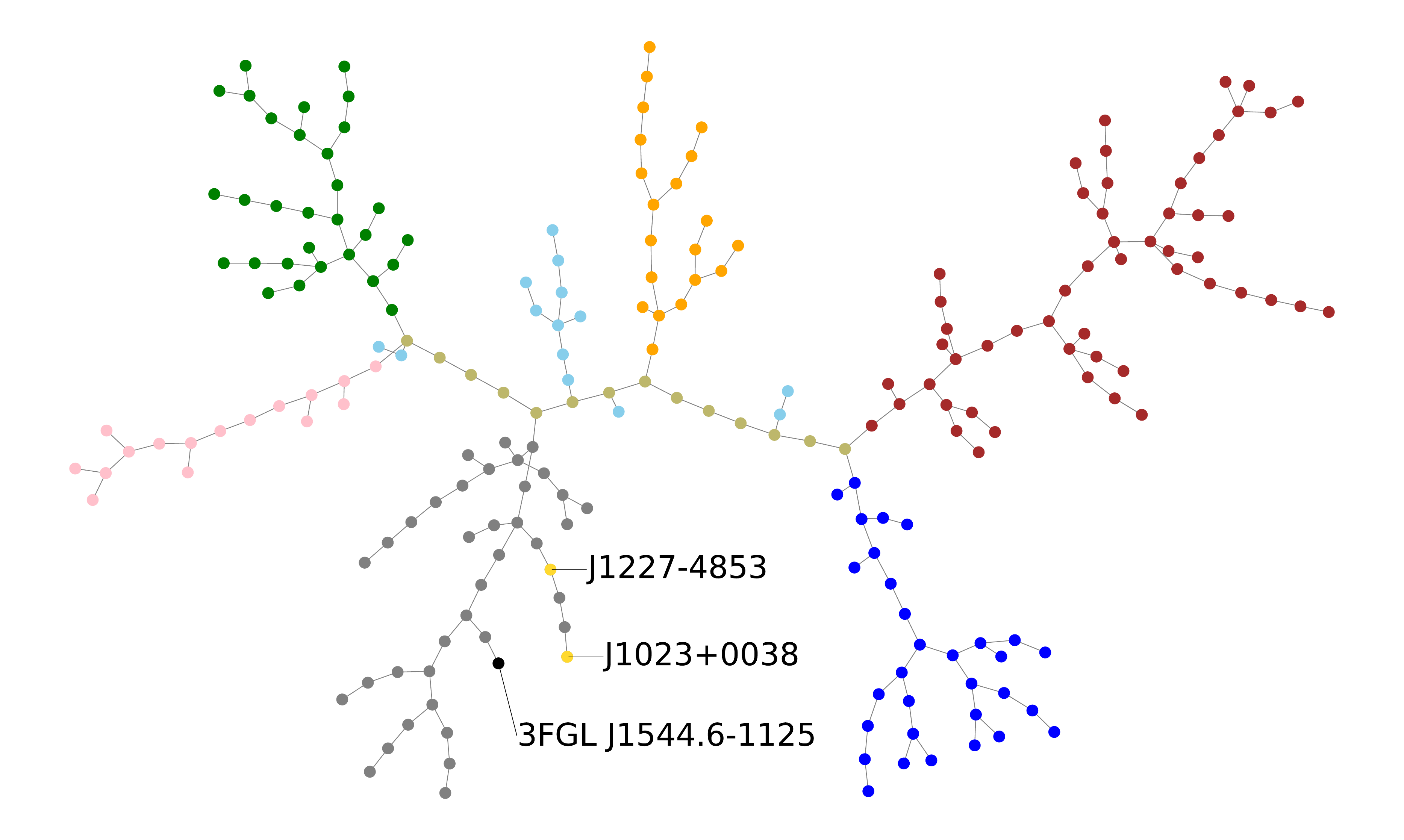}
\caption{The MST defined as $T'(219,218)$ containing the 3FGL J1544.6-1125 (in black), is shown with the known tMSPs (in yellow) labeled, J1023+0038 and J1227-4853, as seen in Fig. \ref{fig: MST_MSP} and the color code of the significant branches, trunk, and the remaining pulsars as seen in Fig. \ref{fig: MST_siginificance_branches}.
}
       \label{fig: MST_3FGL1544}
\end{figure}

We observe that the stability of $T'$ is extremely high concerning $T$, since in more than 99\% of the cases, they only differ on one edge (isomorphism, see e.g. \citealt{valiente:2002}).
As seen in the previous analysis, this allows us to position 3FGL J1544.6-1125 considering known parts of $T$ as the regions with the marked classes of pulsars as in Fig. \ref{fig: MST_MSP} or the branches seen in Fig. \ref{fig: MST_siginificance_branches}.
In 67\% of the cases, the position of 3FGL J1544.6-1125 falls in a branch (gray), where the known tMSPs are located, as we show in Fig. \ref{fig: MST_3FGL1544}.
The MST then promotes the tMSP classification of 3FGL J1544.6-1125 despite the uncertainties of the variables. 

\subsection{SAX J1808.4-3658}

\begin{figure}
\centering
\includegraphics[width=0.5\textwidth]{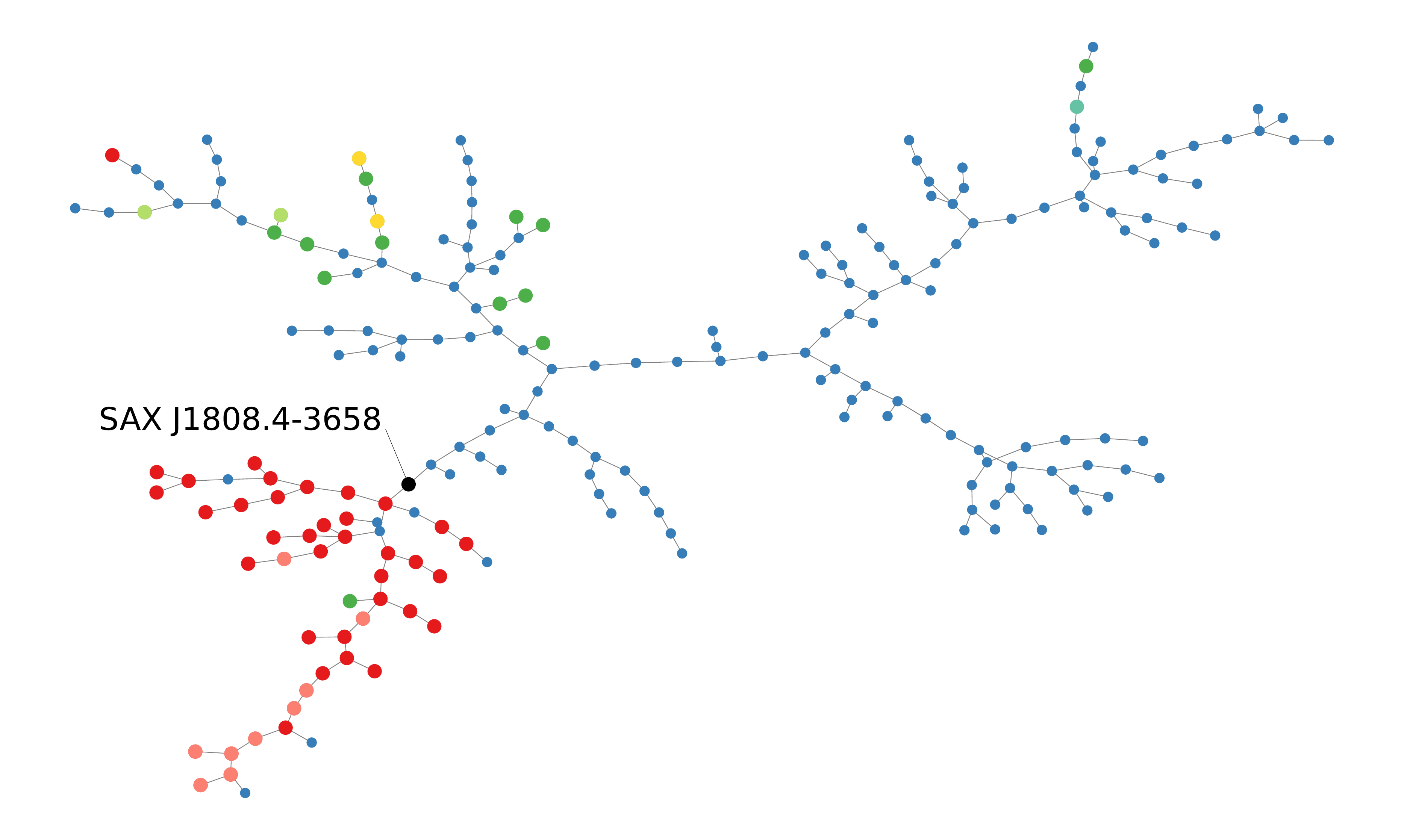}
\caption{The MST defined as $T'(219,218)$, containing the AMXP SAX J1808.4-3658 (in black) is shown. The confirmed BWs appear in red, RBs in green, tMSPs in yellow, the BWs and RBs in globular clusters in light red and light green, the RB candidate in light teal, and keeping the unclassified ones in blue, as seen in Fig. \ref{fig: MST_MSP}.
}
       \label{fig: MST_SAX J1808.4-3658}
\end{figure}

The location of SAX J1808.4-3658, an accreting millisecond X-ray pulsar (AMXP, see eg., \citealt{Patruno_Watts}), within the MST is also of considerable interest. 
It has exhibited ten approximately month-long outbursts with a recurrence of about 2–3 years, making it the AMXP with the most numerous outbursts, suitable for in-depth investigation of its long-term timing properties, as discussed by \cite{Illiano2023}.
SAX J1808.4-3658 is likely a gamma-ray source \citep{deona2016}, and there is an ambiguity about its activity status as a rotation-powered millisecond pulsar during quiescent periods, despite the lack of detected radio pulsations. 
Focusing on the long-term first derivative of the spin frequency, as detailed in Figure 2 and Section 3.2 of \cite{Illiano2023}, reveals a $\dot{\nu} = -1.152(56) \times 10^{-15} \, \mathrm{Hz \, s^{-1}}$,  in alignment with findings from previous works \citep{Patruno_2012, Sanna_2017, Bult_2020}. 
The variables taken are $P = 0.00249391976403(31)$ s, $A_1 = 0.0628033(57)$ lt-s, and the $P_B = 0.083902314(96)$ d (as listed in Table 1 of \citealt{Illiano2023}).
Consequently, we estimate $\dot{P} = - \dot{\nu}/\nu^2 \,  = - P^2 \, \dot{\nu} \sim 7.2 \times 10^{-21} \mathrm{s \, s^{-1}}$.
Past timing analysis indicates that the neutron star orbits a semi-degenerate companion of $M_{C}\sim0.05$ M$_\odot$ (see \citealt{Bildsten_2001}).
Considering the values computed for the other variables, given the known spin period and spin period derivative, we apply the incremental algorithm to obtain $T'$.
Using the BWs and RBs seen in Fig. \ref{fig: MST_MSP} as a reference, we show $T'$ in Fig. \ref{fig: MST_SAX J1808.4-3658} together with the AMXP SAX J1808.4-3658, which falls at the gate of the high-density zone of BWs.

\section{Concluding remarks}
\label{conclusions}

Graph theory provides an elegant way to inspect a population.
A combination of both the MST and the distance ranking underlying helps to distinguish components, as well as to identify candidates for membership in a particular group.
Here we have focused on the MSP population, and have seen that its MST clearly groups. BWs and RBs in different categories. This separation is evident when plotting the $M_C$ for each system.
When the values of $M_C$ are not clear-cut between these populations, it is increasingly difficult to make a safe classification without further studies.
Using the location in the MST, the individual distance ranking for a given pulsar, and the fact that in only very few cases the ranking of a given pulsar thought to pertain to one class is dominated by pulsars belonging to another, suggest that 
\begin{itemize}

\item We promote the BW classification of J1300+1240, J1630+3550, J1317-0157, J1221-0633, J1627+3219, J1701-3006F, and J1737-0314A.

\item The MST location and the ranking favor, or at the very least do not rule out, a BW classification for J1908+2105.

\item The pulsars J1723-2837, J1902-5105, and J1431-4715 are the only ones in the sample containing at least one known
tMSP in their top 3 neighbors. We promote further studies of these pulsars to catch them transitioning into an accretion-powered state in analogy to the known tMSPs.

\item The MST location of IGR J18245-2452/PSR J1824-2452I as an RB transitional pulsar 
implies that the yet unknown $\dot P$ is in the range $\dot{P}\sim (1\times 10^{-20}, 6\times 10^{-19}) \, \mathrm{ss^{-1}}$. Searches limited to this range could empower the techniques and lead to detection.

\item Spanning over the uncertainty in the variables of the putative transitional pulsar behind 3FGL J1544.6-1125, we see indeed that it is located near the others in most of the cases. 

\item The AMXP SAX J1808.4-3658 is close to the group BWs and can be classified as such from its location in the MST. Indeed, \cite{diSalvo2008} proposed that the system is a hidden BW and that in X-ray quiescence the source is ejecting matter. Note that this fact alone would not necessarily imply its possible BW nature. 
    
\item If we use graph theory to cluster the population of MSPs via distinguishing branches (unsupervised methodology), we recover a clear distinction between BWs, RBs, and a significant number of other MSPs showing different features, larger $P_{B}$, $M_{C}$, $P$, and smaller $\dot E$ releases. 
    
\item This process also separates BWs in the field and these in globular clusters, placing them in different branches within the BW region.
    
\end{itemize}

\section*{Acknowledgements}
This work was supported by the grant PID2021-124581OB-I00 of MCIU/AEI/10.13039/501100011033 and 2021SGR00426. 
This work was also supported by the program Unidad de Excelencia María de Maeztu CEX2020-001058-M and by MCIU with funding from the European Union NextGeneration EU (PRTR-C17.I1). 
CRG is funded by the Ph.D. FPI fellowship PRE2019-090828 acknowledges the graduate program of the Universitat Aut\`onoma of Barcelona. 
FCZ is supported by a Ram\'on y Cajal fellowship (grant agreement RYC2021-030888-I).
GI, FCZ, AP, and DdM acknowledge financial support from the Italian National Institute for
Astrophysics (INAF) Research Grant ``Uncovering the optical beat of the fastest magnetized neutron stars'' (FANS; PI: AP).
GI and AP also acknowledge financial support from the Italian Ministry of University and Research (MUR) under PRIN 2020 grant No. 2020BRP57Z ‘Gravitational and Electromagnetic-wave Sources in the Universe with Current and Next-generation detectors (GEMS)’.
GI is supported by the AASS Ph.D. joint research program between the University of Rome “Sapienza” and the University of Rome “Tor Vergata” with the collaboration of the National Institute of Astrophysics (INAF).



\bibliographystyle{aa}
\bibliography{biblio}

\begin{thebibliography}{105}
\expandafter\ifx\csname natexlab\endcsname\relax\def\natexlab#1{#1}\fi

\bibitem[{An {et~al.}(2020)An, Romani, Kerr, \& (Fermi-LAT collaboration)}]{An_2020}
An, H., Romani, R.~W., Kerr, M., \& (Fermi-LAT collaboration). 2020, The Astrophysical Journal, 897, 52

\bibitem[{{Archibald} {et~al.}(2015){Archibald}, {Bogdanov}, {Patruno}, {Hessels}, {Deller}, {Bassa}, {Janssen}, {Kaspi}, {Lyne}, {Stappers}, {Tendulkar}, {D'Angelo}, \& {Wijnands}}]{Archibald_2015ApJ}
{Archibald}, A.~M., {Bogdanov}, S., {Patruno}, A., {et~al.} 2015, \apj, 807, 62

\bibitem[{{Archibald} {et~al.}(2009){Archibald}, {Stairs}, {Ransom}, {Kaspi}, {Kondratiev}, {Lorimer}, {McLaughlin}, {Boyles}, {Hessels}, {Lynch}, {van Leeuwen}, {Roberts}, {Jenet}, {Champion}, {Rosen}, {Barlow}, {Dunlap}, \& {Remillard}}]{Archibald2009}
{Archibald}, A.~M., {Stairs}, I.~H., {Ransom}, S.~M., {et~al.} 2009, Science, 324, 1411

\bibitem[{{Baglio} {et~al.}(2023){Baglio}, {Coti Zelati}, {Campana}, {Busquet}, {D'Avanzo}, {Giarratana}, {Giroletti}, {Ambrosino}, {Crespi}, {Miraval Zanon}, {Hou}, {Li}, {Li}, {Wang}, {Russell}, {Torres}, {Alabarta}, {Casella}, {Covino}, {Bramich}, {de Martino}, {M{\'e}ndez}, {Motta}, {Papitto}, {Saikia}, \& {Vincentelli}}]{Baglio_2023A&A}
{Baglio}, M.~C., {Coti Zelati}, F., {Campana}, S., {et~al.} 2023, \aap, 677, A30

\bibitem[{{Baron} \& {M{\'e}nard}(2021)}]{Baron}
{Baron}, D. \& {M{\'e}nard}, B. 2021, \apj, 916, 91

\bibitem[{{Bassa} {et~al.}(2014){Bassa}, {Patruno}, {Hessels}, {Keane}, {Monard}, {Mahony}, {Bogdanov}, {Corbel}, {Edwards}, {Archibald}, {Janssen}, {Stappers}, \& {Tendulkar}}]{Bassa2014}
{Bassa}, C.~G., {Patruno}, A., {Hessels}, J.~W.~T., {et~al.} 2014, \mnras, 441, 1825

\bibitem[{{Bates} {et~al.}(2015){Bates}, {Thornton}, {Bailes}, {Barr}, {Bassa}, {Bhat}, {Burgay}, {Burke-Spolaor}, {Champion}, {Flynn}, {Jameson}, {Johnston}, {Keith}, {Kramer}, {Levin}, {Lyne}, {Milia}, {Ng}, {Petroff}, {Possenti}, {Stappers}, {van Straten}, \& {Tiburzi}}]{bates2015mnras}
{Bates}, S.~D., {Thornton}, D., {Bailes}, M., {et~al.} 2015, \mnras, 446, 4019

\bibitem[{Bildsten \& Chakrabarty(2001)}]{Bildsten_2001}
Bildsten, L. \& Chakrabarty, D. 2001, The Astrophysical Journal, 557, 292

\bibitem[{{Bogdanov} {et~al.}(2015){Bogdanov}, {Archibald}, {Bassa}, {Deller}, {Halpern}, {Heald}, {Hessels}, {Janssen}, {Lyne}, {Mold{\'o}n}, {Paragi}, {Patruno}, {Perera}, {Stappers}, {Tendulkar}, {D'Angelo}, \& {Wijnands}}]{Bogdanov_2015ApJ}
{Bogdanov}, S., {Archibald}, A.~M., {Bassa}, C., {et~al.} 2015, \apj, 806, 148

\bibitem[{{Bogdanov} {et~al.}(2014){Bogdanov}, {Esposito}, {Crawford}, {Possenti}, {McLaughlin}, \& {Freire}}]{Bogdanov2014}
{Bogdanov}, S., {Esposito}, P., {Crawford}, Fronefield, I., {et~al.} 2014, \apj, 781, 6

\bibitem[{Bogdanov \& Halpern(2015)}]{Bogdanov_2015}
Bogdanov, S. \& Halpern, J.~P. 2015, The Astrophysical Journal Letters, 803, L27

\bibitem[{Braglia {et~al.}(2020)Braglia, Mignani, Belfiore, Marelli, Israel, Novara, De Luca, Tiengo, \& Saz Parkinson}]{Braglia2020}
Braglia, C., Mignani, R.~P., Belfiore, A., {et~al.} 2020, Monthly Notices of the Royal Astronomical Society, 497, 5364

\bibitem[{Brandes(2001)}]{Brandes}
Brandes, U. 2001, The Journal of Mathematical Sociology, 25, 163

\bibitem[{Britt {et~al.}(2017)Britt, Strader, Chomiuk, Tremou, Peacock, Halpern, \& Salinas}]{Britt_2017}
Britt, C.~T., Strader, J., Chomiuk, L., {et~al.} 2017, The Astrophysical Journal, 849, 21

\bibitem[{{Bult} {et~al.}(2020){Bult}, {Chakrabarty}, {Arzoumanian}, {Gendreau}, {Guillot}, {Malacaria}, {Ray}, \& {Strohmayer}}]{Bult_2020}
{Bult}, P., {Chakrabarty}, D., {Arzoumanian}, Z., {et~al.} 2020, \apj, 898, 38

\bibitem[{{Camilo} {et~al.}(2015){Camilo}, {Kerr}, {Ray}, {Ransom}, {Sarkissian}, {Cromartie}, {Johnston}, {Reynolds}, {Wolff}, {Freire}, {Bhattacharyya}, {Ferrara}, {Keith}, {Michelson}, {Saz Parkinson}, \& {Wood}}]{Camilo_2015ApJ}
{Camilo}, F., {Kerr}, M., {Ray}, P.~S., {et~al.} 2015, \apj, 810, 85

\bibitem[{{Cassanelli} {et~al.}(2022){Cassanelli}, {Naletto}, {Codogno}, {Barbieri}, {Verroi}, \& {Zampieri}}]{PCA_application}
{Cassanelli}, T., {Naletto}, G., {Codogno}, G., {et~al.} 2022, \aap, 663, A106

\bibitem[{{Corcoran} {et~al.}(2023){Corcoran}, {Ransom}, \& {Lynch}}]{Corcoran2023}
{Corcoran}, K.~A., {Ransom}, S.~M., \& {Lynch}, R.~S. 2023, Research Notes of the American Astronomical Society, 7, 41

\bibitem[{{Corongiu} {et~al.}(2021){Corongiu}, {Mignani}, {Seyffert}, {Clark}, {Venter}, {Nieder}, {Possenti}, {Burgay}, {Belfiore}, {De Luca}, {Ridolfi}, \& {Wadiasingh}}]{Corongiu_2021}
{Corongiu}, A., {Mignani}, R.~P., {Seyffert}, A.~S., {et~al.} 2021, \mnras, 502, 935

\bibitem[{{Coti Zelati} {et~al.}(2019){Coti Zelati}, {Papitto}, {de Martino}, {Buckley}, {Odendaal}, {Li}, {Russell}, {Torres}, {Mazzola}, {Bozzo}, {Gromadzki}, {Campana}, {Rea}, {Ferrigno}, \& {Migliari}}]{CotiZelati_2019}
{Coti Zelati}, F., {Papitto}, A., {de Martino}, D., {et~al.} 2019, \aap, 622, A211

\bibitem[{{Crawford} {et~al.}(2013){Crawford}, {Lyne}, {Stairs}, {Kaplan}, {McLaughlin}, {Freire}, {Burgay}, {Camilo}, {D'Amico}, {Faulkner}, {Kramer}, {Lorimer}, {Manchester}, {Possenti}, \& {Steeghs}}]{Crawford2013}
{Crawford}, F., {Lyne}, A.~G., {Stairs}, I.~H., {et~al.} 2013, \apj, 776, 20

\bibitem[{{de Amorim} {et~al.}(2022){de Amorim}, {Cavalcanti}, \& {Cruz}}]{RobustScaler}
{de Amorim}, L. B.~V., {Cavalcanti}, G. D.~C., \& {Cruz}, R. M.~O. 2022, arXiv e-prints, arXiv:2212.12343

\bibitem[{{de Martino} {et~al.}(2013){de Martino}, {Belloni}, {Falanga}, {Papitto}, {Motta}, {Pellizzoni}, {Evangelista}, {Piano}, {Masetti}, {Bonnet-Bidaud}, {Mouchet}, {Mukai}, \& {Possenti}}]{deMartino_2013A&A}
{de Martino}, D., {Belloni}, T., {Falanga}, M., {et~al.} 2013, \aap, 550, A89

\bibitem[{{de Martino} {et~al.}(2024){de Martino}, {Phosrisom}, {Dhillon}, {Torres}, {Coti Zelati}, {Breton}, {Marsh}, {Miraval Zanon}, {Rea}, \& {Papitto}}]{deMartino2024}
{de Martino}, D., {Phosrisom}, A., {Dhillon}, V.~S., {et~al.} 2024, arXiv e-prints, arXiv:2409.02075

\bibitem[{{de O{\~n}a Wilhelmi} {et~al.}(2016){de O{\~n}a Wilhelmi}, {Papitto}, {Li}, {Rea}, {Torres}, {Burderi}, {Di Salvo}, {Iaria}, {Riggio}, \& {Sanna}}]{deona2016}
{de O{\~n}a Wilhelmi}, E., {Papitto}, A., {Li}, J., {et~al.} 2016, \mnras, 456, 2647

\bibitem[{{Deller} {et~al.}(2012){Deller}, {Archibald}, {Brisken}, {Chatterjee}, {Janssen}, {Kaspi}, {Lorimer}, {Lyne}, {McLaughlin}, {Ransom}, {Stairs}, \& {Stappers}}]{Deller_2012ApJ}
{Deller}, A.~T., {Archibald}, A.~M., {Brisken}, W.~F., {et~al.} 2012, \apjl, 756, L25

\bibitem[{{Deneva} {et~al.}(2021){Deneva}, {Ray}, {Camilo}, {Freire}, {Cromartie}, {Ransom}, {Ferrara}, {Kerr}, {Burnett}, \& {Parkinson}}]{Deneva_2021}
{Deneva}, J.~S., {Ray}, P.~S., {Camilo}, F., {et~al.} 2021, \apj, 909, 6

\bibitem[{{di Salvo} {et~al.}(2008){di Salvo}, {Burderi}, {Riggio}, {Papitto}, \& {Menna}}]{diSalvo2008}
{di Salvo}, T., {Burderi}, L., {Riggio}, A., {Papitto}, A., \& {Menna}, M.~T. 2008, \mnras, 389, 1851

\bibitem[{{Di Salvo} {et~al.}(2023){Di Salvo}, {Papitto}, {Marino}, {Iaria}, \& {Burderi}}]{DiSalvo2023}
{Di Salvo}, T., {Papitto}, A., {Marino}, A., {Iaria}, R., \& {Burderi}, L. 2023, arXiv e-prints, arXiv:2311.12516

\bibitem[{{Douglas} {et~al.}(2022){Douglas}, {Padmanabh}, {Ransom}, {Ridolfi}, {Freire}, {Krishnan}, {Barr}, {Pallanca}, {Cadelano}, {Possenti}, {Stairs}, {Hessels}, {DeCesar}, {Lynch}, {Bailes}, {Burgay}, {Champion}, {Karuppusamy}, {Kramer}, {Stappers}, \& {Vleeschower}}]{Douglas2022}
{Douglas}, A., {Padmanabh}, P.~V., {Ransom}, S.~M., {et~al.} 2022, \apj, 927, 126

\bibitem[{{Eckert} {et~al.}(2013){Eckert}, {Del Santo}, {Bazzano}, {Watanabe}, {Paizis}, {Bozzo}, {Ferrigno}, {Caballero}, {Sidoli}, \& {Kuiper}}]{IGRJ1824_INTEGRAL}
{Eckert}, D., {Del Santo}, M., {Bazzano}, A., {et~al.} 2013, The Astronomer's Telegram, 4925, 1

\bibitem[{{Eichler} \& {Levinson}(1988)}]{Eichler1988}
{Eichler}, D. \& {Levinson}, A. 1988, \apjl, 335, L67

\bibitem[{{Erickson}(2019)}]{Erickson2019}
{Erickson}, J. 2019, Algorithms (Jeff Erickson, Lecture Notes University of Illinois at Urbana-Champaign)

\bibitem[{{Ferrigno} {et~al.}(2014){Ferrigno}, {Bozzo}, {Papitto}, {Rea}, {Pavan}, {Campana}, {Wieringa}, {Filipovi{\'c}}, {Falanga}, \& {Stella}}]{Ferrigno2014}
{Ferrigno}, C., {Bozzo}, E., {Papitto}, A., {et~al.} 2014, \aap, 567, A77

\bibitem[{Freeman(1977)}]{original}
Freeman, L.~C. 1977, Sociometry, 40, 35

\bibitem[{Freire {et~al.}(2005)Freire, Hessels, Nice, Ransom, Lorimer, \& Stairs}]{Freire_2005}
Freire, P. C.~C., Hessels, J. W.~T., Nice, D.~J., {et~al.} 2005, The Astrophysical Journal, 621, 959

\bibitem[{{Freire} {et~al.}(2017){Freire}, {Ridolfi}, {Kramer}, {Jordan}, {Manchester}, {Torne}, {Sarkissian}, {Heinke}, {D'Amico}, {Camilo}, {Lorimer}, \& {Lyne}}]{Freire2017}
{Freire}, P.~C.~C., {Ridolfi}, A., {Kramer}, M., {et~al.} 2017, \mnras, 471, 857

\bibitem[{{Garc{\'\i}a} \& {Torres}(2023)}]{MST-2}
{Garc{\'\i}a}, C.~R. \& {Torres}, D.~F. 2023, \mnras [\eprint[arXiv]{2301.05408}]

\bibitem[{{Garc{\'\i}a} {et~al.}(2022){Garc{\'\i}a}, {Torres}, \& {Patruno}}]{MST-1}
{Garc{\'\i}a}, C.~R., {Torres}, D.~F., \& {Patruno}, A. 2022, \mnras, 515, 3883

\bibitem[{{Gower} \& {Ross}(1969)}]{Gower1969}
{Gower}, J.~C. \& {Ross}, G. J.~S. 1969, Journal of the Royal Statistical Society. Series C (Applied Statistics), 18, 54

\bibitem[{{Hui}(2014)}]{Hui2014}
{Hui}, C.-Y. 2014, Journal of Astronomy and Space Sciences, 31, 101

\bibitem[{{Illiano} {et~al.}(2023){Illiano}, {Papitto}, {Sanna}, {Bult}, {Ambrosino}, {Miraval Zanon}, {Coti Zelati}, {Stella}, {Altamirano}, {Baglio}, {Bozzo}, {Burderi}, {de Martino}, {Di Marco}, {di Salvo}, {Ferrigno}, {Loktev}, {Marino}, {Ng}, {Pilia}, {Poutanen}, \& {Salmi}}]{Illiano2023}
{Illiano}, G., {Papitto}, A., {Sanna}, A., {et~al.} 2023, \apjl, 942, L40

\bibitem[{{Jiang} {et~al.}(2013){Jiang}, {Zhang}, {Tanni}, \& {Zhao}}]{Jian2013}
{Jiang}, L., {Zhang}, C.-M., {Tanni}, A., \& {Zhao}, H.-H. 2013, in International Journal of Modern Physics Conference Series, Vol.~23, International Journal of Modern Physics Conference Series, 95--98

\bibitem[{Karttunen {et~al.}(2007)Karttunen, Kröger, Oja, Poutanen, \& Donner}]{FundamentalAstronomy}
Karttunen, H., Kröger, P., Oja, H., Poutanen, M., \& Donner, K.~J. 2007, Fundamental Astronomy, 5th edn. (Springer)

\bibitem[{Kennedy {et~al.}(2022)Kennedy, Breton, Clark, Mata Sánchez, Voisin, Dhillon, Halpern, Marsh, Nieder, Ray, \& van Kerkwijk}]{Kennedy_2022}
Kennedy, M.~R., Breton, R.~P., Clark, C.~J., {et~al.} 2022, Monthly Notices of the Royal Astronomical Society, 512, 3001–3014

\bibitem[{{Kerr} {et~al.}(2012){Kerr}, {Camilo}, {Johnson}, {Ferrara}, {Guillemot}, {Harding}, {Hessels}, {Johnston}, {Keith}, {Kramer}, {Ransom}, {Ray}, {Reynolds}, {Sarkissian}, \& {Wood}}]{Kerr_2012ApJ}
{Kerr}, M., {Camilo}, F., {Johnson}, T.~J., {et~al.} 2012, \apjl, 748, L2

\bibitem[{{King} {et~al.}(2005){King}, {Beer}, {Rolfe}, {Schenker}, \& {Skipp}}]{King2005}
{King}, A.~R., {Beer}, M.~E., {Rolfe}, D.~J., {Schenker}, K., \& {Skipp}, J.~M. 2005, \mnras, 358, 1501

\bibitem[{Kiziltan \& Thorsett(2010)}]{Kiziltan_2010}
Kiziltan, B. \& Thorsett, S.~E. 2010, The Astrophysical Journal, 715, 335–341

\bibitem[{{Kleinberg} \& {Tardos}(2005)}]{Kleinberg2005}
{Kleinberg}, J. \& {Tardos}, E. 2005, Algorithm design (Addison Wesley)

\bibitem[{{Koljonen} \& {Linares}(2023)}]{Linares2023}
{Koljonen}, K. I.~I. \& {Linares}, M. 2023, \mnras, 525, 3963

\bibitem[{{Konacki} \& {Wolszczan}(2003)}]{Konacki_2002}
{Konacki}, M. \& {Wolszczan}, A. 2003, \apjl, 591, L147

\bibitem[{{Kong} {et~al.}(2017){Kong}, {Hui}, {Takata}, {Li}, \& {Tam}}]{Kong2017}
{Kong}, A.~K.~H., {Hui}, C.~Y., {Takata}, J., {Li}, K.~L., \& {Tam}, P.~H.~T. 2017, \apj, 839, 130

\bibitem[{Kruskal(1956)}]{Kruskal1956}
Kruskal, J.~B. 1956, Proc. Amer. Math. Soc., 7, 48

\bibitem[{{Lee} {et~al.}(2018){Lee}, {Hui}, {Takata}, {Kong}, {Tam}, \& {Cheng}}]{Lee2018}
{Lee}, J., {Hui}, C.~Y., {Takata}, J., {et~al.} 2018, \apj, 864, 23

\bibitem[{Lehmann(2012)}]{KS-test5}
Lehmann, E.~L. 2012, in Selected Works of E.L. Lehmann (Springer), 373--390

\bibitem[{{Lewis} {et~al.}(2023){Lewis}, {Olszanski}, {Deneva}, {Freire}, {McLaughlin}, {Stovall}, {Bagchi}, {Martinez}, \& {Perera}}]{Lewis_2023}
{Lewis}, E.~F., {Olszanski}, T. E.~E., {Deneva}, J.~S., {et~al.} 2023, \apj, 956, 132

\bibitem[{{Li} {et~al.}(2014){Li}, {Halpern}, \& {Thorstensen}}]{Li2014}
{Li}, M., {Halpern}, J.~P., \& {Thorstensen}, J.~R. 2014, \apj, 795, 115

\bibitem[{{Linares}(2014)}]{Linares2014}
{Linares}, M. 2014, \apj, 795, 72

\bibitem[{{Linares} {et~al.}(2014){Linares}, {Bahramian}, {Heinke}, {Wijnands}, {Patruno}, {Altamirano}, {Homan}, {Bogdanov}, \& {Pooley}}]{Linares_2014MNRAS}
{Linares}, M., {Bahramian}, A., {Heinke}, C., {et~al.} 2014, \mnras, 438, 251

\bibitem[{{Lorimer} \& {Kramer}(2012)}]{Lorimer2012}
{Lorimer}, D.~R. \& {Kramer}, M. 2012, {Handbook of Pulsar Astronomy} (Cambridge University Press)

\bibitem[{Lynch {et~al.}(2012)Lynch, Freire, Ransom, \& Jacoby}]{Lynch_2012}
Lynch, R.~S., Freire, P. C.~C., Ransom, S.~M., \& Jacoby, B.~A. 2012, The Astrophysical Journal, 745, 109

\bibitem[{{Manchester}(2017)}]{Manchester2017}
{Manchester}, R.~N. 2017, Journal of Astrophysics and Astronomy, 38, 42

\bibitem[{{Manchester} {et~al.}(2005){Manchester}, {Hobbs}, {Teoh}, \& {Hobbs}}]{ATNF-Catalog}
{Manchester}, R.~N., {Hobbs}, G.~B., {Teoh}, A., \& {Hobbs}, M. 2005, \aj, 129, 1993

\bibitem[{{Maritz} {et~al.}(2016){Maritz}, {Maritz}, \& {Meintjes}}]{Maritz2016}
{Maritz}, J., {Maritz}, E., \& {Meintjes}, P. 2016, The Proceedings of SAIP2016, the 61st Annual Conference of the South African Institute of Physics, edited by Steve Peterson and Sahal Yacoob (UCT/2016), pp. 243 - 248. ISBN: 978-0-620-77094-1.

\bibitem[{{Miraval Zanon} {et~al.}(2018){Miraval Zanon}, {Burgay}, {Possenti}, \& {Ridolfi}}]{MiravalZanon_2018J}
{Miraval Zanon}, A., {Burgay}, M., {Possenti}, A., \& {Ridolfi}, A. 2018, in Journal of Physics Conference Series, Vol. 956, Journal of Physics Conference Series (IOP), 012004

\bibitem[{Moxley \& Moxley(1974)}]{Moxley1974}
Moxley, R.~L. \& Moxley, N.~F. 1974, Sociometry, 37, 122

\bibitem[{{Papitto} {et~al.}(2019){Papitto}, {Ambrosino}, {Stella}, {Torres}, {Coti Zelati}, {Ghedina}, {Meddi}, {Sanna}, {Casella}, {Dallilar}, {Eikenberry}, {Israel}, {Onori}, {Piranomonte}, {Bozzo}, {Burderi}, {Campana}, {de Martino}, {Di Salvo}, {Ferrigno}, {Rea}, {Riggio}, {Serrano}, {Veledina}, \& {Zampieri}}]{Papitto_2019ApJ}
{Papitto}, A., {Ambrosino}, F., {Stella}, L., {et~al.} 2019, \apj, 882, 104

\bibitem[{{Papitto} \& {Bhattacharyya}(2022)}]{Papitto2022}
{Papitto}, A. \& {Bhattacharyya}, S. 2022, {Millisecond pulsars} (Springer ASSL)

\bibitem[{{Papitto} \& {de Martino}(2022)}]{Papitto_tMSPSectionBook}
{Papitto}, A. \& {de Martino}, D. 2022, in Astrophysics and Space Science Library, Vol. 465, Astrophysics and Space Science Library, ed. S.~{Bhattacharyya}, A.~{Papitto}, \& D.~{Bhattacharya}, 157--200

\bibitem[{Papitto {et~al.}(2013)Papitto, Ferrigno, Bozzo, Rea, Pavan, Burderi, Burgay, Campana, Di~Salvo, Falanga, Filipović, Freire, Hessels, Possenti, Ransom, Riggio, Romano, Sarkissian, Stairs, Stella, Torres, Wieringa, \& Wong}]{Papitto2013}
Papitto, A., Ferrigno, C., Bozzo, E., {et~al.} 2013, nature, 501, 517

\bibitem[{{Patruno} {et~al.}(2014){Patruno}, {Archibald}, {Hessels}, {Bogdanov}, {Stappers}, {Bassa}, {Janssen}, {Kaspi}, {Tendulkar}, \& {Lyne}}]{Patruno2014}
{Patruno}, A., {Archibald}, A.~M., {Hessels}, J.~W.~T., {et~al.} 2014, \apjl, 781, L3

\bibitem[{{Patruno} {et~al.}(2012){Patruno}, {Bult}, {Gopakumar}, {Hartman}, {Wijnands}, {van der Klis}, \& {Chakrabarty}}]{Patruno_2012}
{Patruno}, A., {Bult}, P., {Gopakumar}, A., {et~al.} 2012, \apjl, 746, L27

\bibitem[{{Patruno} \& {Watts}(2021)}]{Patruno_Watts}
{Patruno}, A. \& {Watts}, A.~L. 2021, in Astrophysics and Space Science Library, Vol. 461, Timing Neutron Stars: Pulsations, Oscillations and Explosions, ed. T.~M. {Belloni}, M.~{M{\'e}ndez}, \& C.~{Zhang}, 143--208

\bibitem[{{Pearson}(1901)}]{Pearson1901}
{Pearson}, K. 1901, The London, Edinburgh, and Dublin Philosophical Magazine and Journal of Science, 2, 559

\bibitem[{{Prim}(1957)}]{Prim1957}
{Prim}, R.~C. 1957, Bell System Technical Journal, 36, 1389

\bibitem[{{Ray} {et~al.}(2012){Ray}, {Abdo}, {Parent}, {Bhattacharya}, {Bhattacharyya}, {Camilo}, {Cognard}, {Theureau}, {Ferrara}, {Harding}, {Thompson}, {Freire}, {Guillemot}, {Gupta}, {Roy}, {Hessels}, {Johnston}, {Keith}, {Shannon}, {Kerr}, {Michelson}, {Romani}, {Kramer}, {McLaughlin}, {Ransom}, {Roberts}, {Saz Parkinson}, {Ziegler}, {Smith}, {Stappers}, {Weltevrede}, \& {Wood}}]{Ray2012}
{Ray}, P.~S., {Abdo}, A.~A., {Parent}, D., {et~al.} 2012, arXiv e-prints, arXiv:1205.3089

\bibitem[{{Roberts}(2013)}]{Roberts2013}
{Roberts}, M. S.~E. 2013, in Neutron Stars and Pulsars: Challenges and Opportunities after 80 years, ed. J.~{van Leeuwen}, Vol. 291, 127--132

\bibitem[{{Roberts} {et~al.}(2018){Roberts}, {Al Noori}, {Torres}, {McLaughlin}, {Gentile}, {Hessels}, {Ransom}, {Ray}, {Kerr}, \& {Breton}}]{Roberts20172018}
{Roberts}, M. S.~E., {Al Noori}, H., {Torres}, R.~A., {et~al.} 2018, in Pulsar Astrophysics the Next Fifty Years, ed. P.~{Weltevrede}, B.~B.~P. {Perera}, L.~L. {Preston}, \& S.~{Sanidas}, Vol. 337, 43--46

\bibitem[{{Roughgarden}(2019)}]{Roughgarden2019}
{Roughgarden}, T. 2019, {Algorithms Illuminated (Part 3): Greedy Algorithms and Dynamic Programming } (Soundlikeyourself Publishing, LLC)

\bibitem[{{Sanna} {et~al.}(2017){Sanna}, {Di Salvo}, {Burderi}, {Riggio}, {Pintore}, {Gambino}, {Iaria}, {Tailo}, {Scarano}, \& {Papitto}}]{Sanna_2017}
{Sanna}, A., {Di Salvo}, T., {Burderi}, L., {et~al.} 2017, \mnras, 471, 463

\bibitem[{{Sen} {et~al.}(2024){Sen}, {Linares}, {Kennedy}, {Breton}, {Misra}, {Turchetta}, {Dhillon}, {Mata Sanchez}, \& {Clark}}]{Sen_2024}
{Sen}, B., {Linares}, M., {Kennedy}, M.~R., {et~al.} 2024, arXiv e-prints, arXiv:2407.10800

\bibitem[{{Shlens}(2014)}]{Shlens2014}
{Shlens}, J. 2014, arXiv e-prints, arXiv:1404.1100

\bibitem[{Smith {et~al.}(2023)Smith, Abdollahi, Ajello, Bailes, Baldini, Ballet, Baring, Bassa, Gonzalez, Bellazzini, Berretta, Bhattacharyya, Bissaldi, Bonino, Bottacini, Bregeon, Bruel, Burgay, Burnett, Cameron, Camilo, Caputo, Caraveo, Cavazzuti, Chiaro, Ciprini, Clark, Cognard, Corongiu, Orestano, Crnogorcevic, Cuoco, Cutini, D’Ammando, de~Angelis, DeCesar, Gaetano, de~Menezes, Deneva, de~Palma, Lalla, Dirirsa, Venere, Domínguez, Dumora, Fegan, Ferrara, Fiori, Fleischhack, Flynn, Franckowiak, Freire, Fukazawa, Fusco, Galanti, Gammaldi, Gargano, Gasparrini, Giacchino, Giglietto, Giordano, Giroletti, Green, Grenier, Guillemot, Guiriec, Gustafsson, Harding, Hays, Hewitt, Horan, Hou, Jankowski, Johnson, Johnson, Johnston, Kataoka, Keith, Kerr, Kramer, Kuss, Latronico, Lee, Li, Li, Limyansky, Longo, Loparco, Lorusso, Lovellette, Lower, Lubrano, Lyne, Maan, Maldera, Manchester, Manfreda, Marelli, Martí-Devesa, Mazziotta, McEnery, Mereu, Michelson, Mickaliger, Mitthumsiri, Mizuno, Moiseev, Monzani, Morselli,
  Negro, Nemmen, Nieder, Nuss, Omodei, Orienti, Orlando, Ormes, Palatiello, Paneque, Panzarini, Parthasarathy, Persic, Pesce-Rollins, Pillera, Poon, Porter, Possenti, Principe, Rainò, Rando, Ransom, Ray, Razzano, Razzaque, Reimer, Reimer, Renault-Tinacci, Romani, Sánchez-Conde, Parkinson, Scotton, Serini, Sgrò, Shannon, Sharma, Shen, Siskind, Spandre, Spinelli, Stappers, Stephens, Suson, Tabassum, Tajima, Tak, Theureau, Thompson, Tibolla, Torres, Valverde, Venter, Wadiasingh, Wang, Wang, Wang, Weltevrede, Wood, Yan, Zaharijas, Zhang, \& Zhu}]{Fermi3PC}
Smith, D.~A., Abdollahi, S., Ajello, M., {et~al.} 2023, The Astrophysical Journal, 958, 191

\bibitem[{Sobey {et~al.}(2022)Sobey, Bassa, O’Sullivan, Callingham, Tan, Hessels, Kondratiev, Stappers, Tiburzi, Heald, Shimwell, Breton, Kirwan, Vedantham, Carretti, Grießmeier, Haverkorn, \& Karastergiou}]{Sobey_2022}
Sobey, C., Bassa, C.~G., O’Sullivan, S.~P., {et~al.} 2022, Astronomy \& Astrophysics, 661, A87

\bibitem[{{Strader} {et~al.}(2019){Strader}, {Swihart}, {Chomiuk}, {Bahramian}, {Britt}, {Cheung}, {Dage}, {Halpern}, {Li}, {Mignani}, {Orosz}, {Peacock}, {Salinas}, {Shishkovsky}, \& {Tremou}}]{strader_redbacks}
{Strader}, J., {Swihart}, S., {Chomiuk}, L., {et~al.} 2019, \apj, 872, 42

\bibitem[{Strader {et~al.}(2019)Strader, Swihart, Chomiuk, Bahramian, Britt, Cheung, Dage, Halpern, Li, Mignani, Orosz, Peacock, Salinas, Shishkovsky, \& Tremou}]{Strader_2019}
Strader, J., Swihart, S., Chomiuk, L., {et~al.} 2019, The Astrophysical Journal, 872, 42

\bibitem[{{Str{\"u}der} {et~al.}(2001){Str{\"u}der}, {Briel}, {Dennerl}, {Hartmann}, {Kendziorra}, {Meidinger}, {Pfeffermann}, {Reppin}, {Aschenbach}, {Bornemann}, {Br{\"a}uninger}, {Burkert}, {Elender}, {Freyberg}, {Haberl}, {Hartner}, {Heuschmann}, {Hippmann}, {Kastelic}, {Kemmer}, {Kettenring}, {Kink}, {Krause}, {M{\"u}ller}, {Oppitz}, {Pietsch}, {Popp}, {Predehl}, {Read}, {Stephan}, {St{\"o}tter}, {Tr{\"u}mper}, {Holl}, {Kemmer}, {Soltau}, {St{\"o}tter}, {Weber}, {Weichert}, {von Zanthier}, {Carathanassis}, {Lutz}, {Richter}, {Solc}, {B{\"o}ttcher}, {Kuster}, {Staubert}, {Abbey}, {Holland}, {Turner}, {Balasini}, {Bignami}, {La Palombara}, {Villa}, {Buttler}, {Gianini}, {Lain{\'e}}, {Lumb}, \& {Dhez}}]{Struder_2001A&A}
{Str{\"u}der}, L., {Briel}, U., {Dennerl}, K., {et~al.} 2001, \aap, 365, L18

\bibitem[{Swiggum {et~al.}(2023)Swiggum, Pleunis, Parent, Kaplan, McLaughlin, Stairs, Spiewak, Agazie, Chawla, DeCesar, Dolch, Fiore, Fonseca, Istrate, Kaspi, Kondratiev, van Leeuwen, Levin, Lewis, Lynch, McEwen, Al~Noori, Ransom, Siemens, \& Surnis}]{Swiggum_2023}
Swiggum, J.~K., Pleunis, Z., Parent, E., {et~al.} 2023, The Astrophysical Journal, 944, 154

\bibitem[{{Swihart} {et~al.}(2022){Swihart}, {Strader}, {Chomiuk}, {Aydi}, {Sokolovsky}, {Ray}, \& {Kerr}}]{swihart_blacwidows}
{Swihart}, S.~J., {Strader}, J., {Chomiuk}, L., {et~al.} 2022, \apj, 941, 199

\bibitem[{Tauris {et~al.}(2012)Tauris, Kramer, \& Langer}]{Tauris_2012}
Tauris, T.~M., Kramer, M., \& Langer, N. 2012, Proceedings of the International Astronomical Union, 8, 137–140

\bibitem[{{Tauris} \& {van den Heuvel}(2006)}]{tauris2003formation}
{Tauris}, T.~M. \& {van den Heuvel}, E.~P.~J. 2006, in Compact stellar X-ray sources, Vol.~39, 623--665

\bibitem[{{Torres} {et~al.}(2017){Torres}, {Ji}, {Li}, {Papitto}, {Rea}, {de O{\~n}a Wilhelmi}, \& {Zhang}}]{Torres2017}
{Torres}, D.~F., {Ji}, L., {Li}, J., {et~al.} 2017, \apj, 836, 68

\bibitem[{{Turner} {et~al.}(2001){Turner}, {Abbey}, {Arnaud}, {Balasini}, {Barbera}, {Belsole}, {Bennie}, {Bernard}, {Bignami}, {Boer}, {Briel}, {Butler}, {Cara}, {Chabaud}, {Cole}, {Collura}, {Conte}, {Cros}, {Denby}, {Dhez}, {Di Coco}, {Dowson}, {Ferrando}, {Ghizzardi}, {Gianotti}, {Goodall}, {Gretton}, {Griffiths}, {Hainaut}, {Hochedez}, {Holland}, {Jourdain}, {Kendziorra}, {Lagostina}, {Laine}, {La Palombara}, {Lortholary}, {Lumb}, {Marty}, {Molendi}, {Pigot}, {Poindron}, {Pounds}, {Reeves}, {Reppin}, {Rothenflug}, {Salvetat}, {Sauvageot}, {Schmitt}, {Sembay}, {Short}, {Spragg}, {Stephen}, {Str{\"u}der}, {Tiengo}, {Trifoglio}, {Tr{\"u}mper}, {Vercellone}, {Vigroux}, {Villa}, {Ward}, {Whitehead}, \& {Zonca}}]{Turner_2001A&A}
{Turner}, M.~J.~L., {Abbey}, A., {Arnaud}, M., {et~al.} 2001, \aap, 365, L27

\bibitem[{Valiente(2002)}]{valiente:2002}
Valiente, G. 2002, Algorithms on Trees and Graphs (Berlin: Springer-Verlag)

\bibitem[{{Vleeschower} {et~al.}(2024){Vleeschower}, {Corongiu}, {Stappers}, {Freire}, {Ridolfi}, {Abbate}, {Ransom}, {Possenti}, {Padmanabh}, {Balakrishnan}, {Kramer}, {Venkatraman Krishnan}, {Zhang}, {Bailes}, {Barr}, {Buchner}, \& {Chen}}]{Vleeschower_2024}
{Vleeschower}, L., {Corongiu}, A., {Stappers}, B.~W., {et~al.} 2024, \mnras, 530, 1436

\bibitem[{{Vohl} {et~al.}(2023){Vohl}, {van Leeuwen}, \& {Maan}}]{Vohl2023}
{Vohl}, D., {van Leeuwen}, J., \& {Maan}, Y. 2023, arXiv e-prints, arXiv:2311.09201

\bibitem[{{Wilson}(2010)}]{Wilson2010}
{Wilson}, R.~J. 2010, Introduction to graph theory (Pearson)

\bibitem[{Wolfe(2012)}]{KS-test4}
Wolfe, D.~A. 2012, in Selected Works of E.L. Lehmann, ed. J.~Rojo (Boston, MA: Springer US), 1101--1110

\bibitem[{{Wolszczan}(1994)}]{Wolszczan_1994}
{Wolszczan}, A. 1994, Science, 264, 538

\bibitem[{{Wolszczan} \& {Frail}(1992)}]{Wolszczan_1992}
{Wolszczan}, A. \& {Frail}, D.~A. 1992, \nat, 355, 145

\bibitem[{{Wolszczan} {et~al.}(2000){Wolszczan}, {Hoffman}, {Konacki}, {Anderson}, \& {Xilouris}}]{Wolszczan_2000}
{Wolszczan}, A., {Hoffman}, I.~M., {Konacki}, M., {Anderson}, S.~B., \& {Xilouris}, K.~M. 2000, \apjl, 540, L41

\bibitem[{Yadolah(2008)}]{KS-test3}
Yadolah, D. 2008, in The Concise Encyclopedia of Statistics (New York, NY: Springer New York), 283--287

\bibitem[{Yan {et~al.}(2013)Yan, Shen, Yuan, Wang, Rottmann, \& Alef}]{Yan2013}
Yan, Z., Shen, Z.-Q., Yuan, J.-P., {et~al.} 2013, Monthly Notices of the Royal Astronomical Society, 433, 162

\bibitem[{{Yap} {et~al.}(2023){Yap}, {Kong}, \& {Li}}]{yap2023light}
{Yap}, Y.~X.~J., {Kong}, A. K.~H., \& {Li}, K.-L. 2023, \apj, 955, 21

\bibitem[{Zhao \& Heinke(2022)}]{Zhao_2022}
Zhao, J. \& Heinke, C.~O. 2022, Monthly Notices of the Royal Astronomical Society, 511, 5964–5983

\end{thebibliography}

\begin{appendix}
\section{Significant branches
\label{sec: appendix1}}

We apply the betweenness centrality estimator (see \citealt{original, Moxley1974, Brandes, Baron}) to identify the most central nodes of the graph $T$.
Values seen as outliers in betweenness centrality will be named potential trunk nodes, or pTNs.
The trunk is defined as a path, i.e., a sequence of non-repeated nodes. 
At the end of the trunk, the pTNs that delimit are called contour nodes (CN).
These CNs must be nodes of a degree greater than 2, so they act as an articulation point for the graph, i.e., removing these nodes partitions the graph into connected components.
The connected components originating in a CN are called contour branches (CB).
This leaves a clustered structure with a minimum of two groupings at each extreme of the trunk. 
To avoid limiting the trunk in nodes of degree 2 or larger but from which the departing structures are of non-representative size (i.e., we aim to avoid noise near the trunk), we define a significant threshold ($\alpha_{s}$) so that the number of nodes in the CBs is requested to exceed this lower limit. 
We set the threshold at 5\%, i.e. implying that branches will contain at least 11 pulsars of the total population.
Once all the possible CNs are identified, we calculate all the possible trunks as the paths that result in pairing these CNs.
The number of trunks obtained, therefore, are
$\# Trunks = \mathrm{CNs} \times (\mathrm{CNs}-1)/2
~.\label{eq: number_trunks}
$
For each possible trunk, we have a set of branches constituted by the connected components starting from every node of a degree larger than 2 (including the CBs) and having $\alpha_{s}$ as a lower limit to the number of nodes each one must contain.

This results in six sets of branches and trunks.
We choose the set that provides greater uniformity in the size of the resulting groups, which contributes to the robustness
of a comparison analysis, such as a non-parametric test such as the Kolmogorov-Smirnov (KS) statistics \citep[see e.g.,][]{KS-test4, KS-test5, KS-test3}. 
In this case, the null hypothesis ($H_0$) is that the distribution of the variables of the nodes of two given
branches is consistent with them coming from the same parent distribution. 
Rejecting this null hypothesis, say at the 95\% confidence level (CL) or better, would lead us to think that two branches may be formed by different pulsars or at different evolution stages.
Table \ref{tab: significance_branches} shows the KS test results for these branches, where the $H_0$ column indicates which branches differ the most under the above assumptions.
It is observed that most of the branches are distinguished for most of the individual variables.

\begin{table}
\scriptsize
\caption{
Comparison (see \S \ref{significance_branches}) of the branches seen in Fig. \ref{fig: MST_siginificance_branches}.
We denote with a number 1 when two branches, according to the KS-test for a 95\% CL, reject $H_0$ (defined in Appendix \ref{sec: appendix1}) for the analyzed variable.
Instead, we denote with a number 0 the opposite case, where we can not reject $H_0$.
The last column shows the accumulated sum for each confrontation between branches, counting for which variables these branches reject $H_0$.
We denote with '*' those colors in which dark is suppressed, alluding to the dark green and dark blue branches, for visualization reasons.
}
\setlength{\tabcolsep}{5pt}
\begin{tabular}{l rrrrrrrrrr | r}
\hline
Branches    &  $B_{lc}$ &  $B_{s}$ &  $\dot{E}$ &  $\eta_{GJ}$ &  $P$ &  $\dot{P}$ &  $\Delta\Phi$ &  $P_{B}$ &   $A_{1}$ & $M_C$ & $H_{0}$ \\
\hline
Green* - Gray & 0 & 1 & 1 & 1 & 0 & 1 & 1 & 1 & 1 & 1 & 8 \\
Green* - Pink & 1 & 1 & 1 & 1 & 0 & 1 & 1 & 0 & 0 & 0 & 6 \\
Green* - Blue* & 1 & 1 & 1 & 0 & 1 & 1 & 1 & 1 & 1 & 1 & 9 \\
Green* - Orange & 0 & 1 & 1 & 1 & 1 & 1 & 1 & 1 & 1 & 1 & 9 \\
Green* - Brown & 1 & 1 & 1 & 1 & 1 & 0 & 1 & 1 & 1 & 1 & 9 \\
Gray - Pink & 1 & 0 & 1 & 1 & 0 & 0 & 1 & 1 & 1 & 1 & 7 \\
Gray - Blue* & 1 & 1 & 1 & 1 & 1 & 1 & 1 & 1 & 1 & 0 & 9 \\
Gray - Orange & 1 & 1 & 1 & 1 & 1 & 1 & 1 & 0 & 0 & 0 & 7 \\
Gray - Brown & 1 & 1 & 1 & 1 & 1 & 1 & 1 & 1 & 1 & 1 & 10 \\
Pink - Blue* & 1 & 0 & 1 & 1 & 1 & 1 & 1 & 1 & 1 & 1 & 9 \\
Pink - Orange & 1 & 1 & 1 & 1 & 0 & 1 & 1 & 1 & 1 & 1 & 9 \\
Pink - Brown & 1 & 1 & 1 & 1 & 1 & 1 & 1 & 1 & 1 & 1 & 10 \\
Blue* - Orange & 1 & 1 & 1 & 1 & 1 & 1 & 1 & 0 & 0 & 0 & 7 \\
Blue* - Brown & 0 & 1 & 0 & 1 & 1 & 1 & 0 & 1 & 0 & 0 & 5 \\
Orange - Brown & 1 & 1 & 1 & 0 & 1 & 1 & 1 & 1 & 1 & 1 & 9 \\
\hline
\end{tabular}
\label{tab: significance_branches}
\end{table}

\section{No moding in identified tMSP
candidates in the radio state
\label{sec: appendix2}}

\begin{figure*}
\centering
\includegraphics[width=0.8\textwidth]{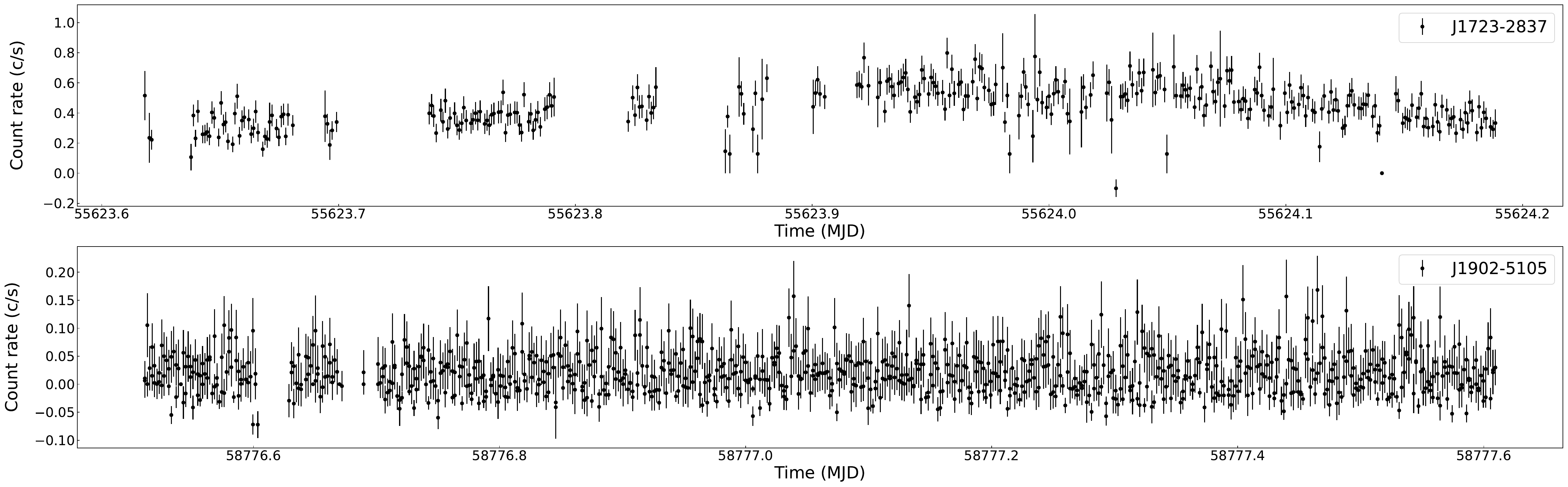}
\caption{\textit{XMM-Newton} 0.3-10 keV background-subtracted light curves over the time interval in which the three EPIC instruments collected data simultaneously for J1723-2837 (top panel) and J1902-5105 (bottom panel). The light curves are binned with a time resolution of 100 s. For a detailed analysis of J1431-4715 see \cite{deMartino2024, Linares2023}.} \label{fig: XMM_light_curves}
\end{figure*}

The EPIC-pn \citep{Struder_2001A&A} operated with a time resolution of 73.4 $\mathrm{ms}$ (full frame mode) for 1723-2837 and J1902-5105, and 47.4 $\mathrm{ms}$ (large window mode) for J1431-4715. The two EPIC-MOS \citep{Turner_2001A&A} were employed with a time resolution of 2.6 s (full frame mode) in the first two cases and 0.9 s (large window mode) for J1431-4715.
We processed and analyzed the data using the Science Analysis Software (SAS; v.21.0.0). 
All three observations exhibited high background activity in the 10–12 keV light curve, so we excluded the contaminated data intervals.
For the EPIC-pn and each MOS, we extracted source photons within a circular region centered on the source position with a 40\arcsec radius, and background photons from an 80\arcsec wide, source-free circular region.
Using the \texttt{epiclccorr} task, 0.3-10 keV background-subtracted light curves were extracted from the three EPIC instruments over the simultaneous coverage time interval, binned with a time resolution of 100 s. 
As previously noted, we recall that the X-ray flux of J1723-2837 (top panel of Fig. \ref{fig: XMM_light_curves}) is variable with the orbital phase as discussed in \citet{Bogdanov2014}.

We searched the \textit{XMM-Newton} light curves for the typical bimodality between 'high' and 'low' intensity modes (see Fig. \ref{fig: XMM_light_curves}; see \cite{deMartino2024}, see also, \citep{Linares2023} for a detailed analysis of J1431-4715). We experimented with different bin sizes but did not observe the expected bimodal feature. Further analysis of the count rate distributions confirmed the absence of this bimodality. A deep search for pulsation has identified PSR J1431-4715 as a gamma-ray pulsar \cite{Fermi3PC}; its gamma-ray emission shows variable signature as those found for prototypical tMSPs \cite{Torres2017}.

Note that the prototype for tMSPs, J1023+0038, is located at $\sim$1.37 kpc \citep{Deller_2012ApJ}, while the estimated distances for J1723-2837 and J1902–5105 are $\sim$0.75 kpc \citep{Bogdanov2014} and $\sim$1.2 kpc \citep{Camilo_2015ApJ}, respectively. Assuming comparable luminosities, if a similar bimodal behavior were present, we would expect to observe it in these closer sources.

Moreover, distinct high and low modes are evident in the faint candidate tMSP 3FGL J1544.6-1125 \citep{Bogdanov_2015}. For the latter source, as a check, we have re-analyzed the \textit{XMM-Newton} archival observation ObsID 072080101. The bimodal pattern was detected in the EPIC/MOS1 0.3-10 keV background-subtracted light curve binned with a time resolution of 100 s and exhibiting an average count rate comparable to that of J1723-2837. Thus, should it be present in J1723-2837, it should have become similarly apparent. 

\end{appendix}

   \end{document}